\documentclass{bmvc2k}

\usepackage{nicefrac}
\usepackage[capitalize]{cleveref}
\usepackage{enumitem}
\usepackage{amssymb}
\usepackage{multirow}

\definecolor{goodgreen}{RGB}{0,115,70}
\definecolor{badred}{RGB}{185,45,45}

\title{Memory Bandwidth, Not FLOPs: Profiling and Accelerating Query-Based Segmentation Decoders on Edge GPUs}

\addauthor{Oliver Busemann}{obusemann@uni-osnabrueck.de}{1,2}
\addauthor{Torben T\"oniges}{Torben.Toeniges@claas.com}{2}
\addauthor{Martin Atzmueller}{martin.atzmueller@uni-osnabrueck.de}{3,4}

\addinstitution{%
 Osnabr\"uck University\\
 Osnabr\"uck, Germany%
}
\addinstitution{%
 CLAAS E-Systems GmbH\\
 Dissen a.T.W., Germany%
}
\addinstitution{%
 Semantic Information Systems Group\\
 Osnabr\"uck University\\
 Osnabr\"uck, Germany%
}
\addinstitution{%
 German Research Center for\\
 Artificial Intelligence (DFKI)\\
 Osnabr\"uck, Germany%
}

\runninghead{Busemann, T\"oniges, Atzmueller}{Memory wall of transformer decoders}

\def\etal{\emph{et al}\bmvaOneDot}

\begin{document}
\maketitle

\begin{abstract}

Query-based segmentation models achieve strong performance even under
real-time inference on workstation GPUs. On edge devices, however, we
show their transformer decoders are deeply memory-bandwidth-bound, a
bottleneck invisible to standard FLOP analysis that worsens with input
resolution and limited DRAM bandwidth. An FP16 roofline analysis with
tensor-core GEMMs active and hardware DRAM counters shows that the
non-tensor-core aggregates of FastInst and Mask2Former reach only
$2.5\%$ to $6\%$ of the CUDA-core ridge yet account for $53\%$ to $70\%$
of decoder time and $61\%$ to $71\%$ of decoder bytes. The largest
individual kernels sustain up to $90\%$ of nominal peak DRAM bandwidth,
effectively saturating the memory bus. The bottleneck is therefore not
primarily low GEMM intensity, but cross-kernel activation traffic.

CompactInst addresses this bottleneck using parameter-free spatial
pooling at $4{\times}$, which reduces the bytes moved within the
transformer decoder, while a two-convolution refinement head recovers
spatial detail. Under TensorRT FP16 on Jetson
Orin~Nano, CompactInst delivers decoder speedups from
$2.35{\times}$ to $4.74{\times}$ and model speedups from
$1.57{\times}$ to $2.62{\times}$ as input resolution grows,
while closely matching accuracy ($-0.4$\,AP on COCO~\texttt{val2017},
$+0.2$\,AP on Cityscapes~\texttt{test}). A heavier variant with
$2.57{\times}$ more decoder FLOPs still runs its decoder $1.74{\times}$
faster on the most bandwidth-constrained device, confirming at the
deployment level that decoder bytes, not FLOPs, govern latency.
\end{abstract}

\section{Introduction}
\label{sec:intro}

Query-based models, which build on the powerful transformer
architecture~\cite{Transformer}, have shown strong performance on
instance segmentation tasks~\cite{Mask2Former}.
FastInst~\cite{FastInst} further demonstrated that such models can
reach real-time inference on high-bandwidth workstation GPUs
(V100, $900$\,GB/s).

While other real-time instance segmentation
methods~\cite{SOLOv2,CondInst,SparseInst} achieve competitive
trade-offs between speed and accuracy, query-based architectures offer
a distinct advantage: the same model can be readily extended to panoptic
segmentation or anomaly detection~\cite{Mask2Anomaly,RbA} with
minimal architectural changes.
This versatility is especially relevant in autonomous driving and
robotics, where multiple perception tasks must run concurrently on a
single edge platform.
Such deployments routinely process inputs at HD or Full~HD
resolution, well beyond the $640{\times}480$ images common in
benchmarks like COCO~\cite{COCO}. This substantially increases the
computational load.
Standard automotive datasets confirm this gap:
Cityscapes~\cite{Cordts2016Cityscapes} captures at $2.1$\,MP,
nuScenes~\cite{nuscenes2019} at $1.4$\,MP,
Waymo~\cite{Sun_2020_CVPR} at $2.5$\,MP, and
BDD100K~\cite{Yu_2020_CVPR} at $0.9$\,MP. These resolutions contain
$3{\times}$ to $8{\times}$ as many pixels as common COCO crops.

However, deploying these models on edge devices such as the NVIDIA
Jetson AGX Orin or Jetson Orin Nano remains challenging due to
significantly higher inference latency compared to workstation hardware.
FLOPs are an unreliable proxy for latency: memory access cost and
degree of parallelism can dominate runtime even when arithmetic
complexity is low~\cite{ShuffleNetV2}, and the widening gap between
compute and memory-bandwidth scaling~\cite{MemoryWall} makes this
mismatch increasingly acute on edge hardware.
Through systematic latency profiling across three platforms spanning a
$6.6{\times}$ range in DRAM bandwidth and a kernel-level FP16 roofline
analysis using measured DRAM traffic on the RTX~A4000, we show that
transformer-decoder latency in the standard FastInst and Mask2Former
decoders is dominated by CUDA-only kernels that stream pixel-token
activations between tensor-core GEMMs. Across the two decoders, the
CUDA-only aggregates reach only $2.5\%$ to $6\%$ of the CUDA ridge yet
account for $53\%$ to $70\%$ of decoder time and $61\%$ to $71\%$ of
DRAM bytes. The largest individual kernels sustain up to $90\%$ of
nominal peak DRAM bandwidth, effectively saturating the memory bus.
The bottleneck is less apparent at FP32 because tensor-core acceleration
at FP16 reduces GEMM time and exposes the non-GEMM activation traffic
(\cref{fig:roofline}).
In FastInst, the pixel decoder produces features at $\nicefrac{1}{8}$
input resolution, yielding $\nicefrac{H}{8} \times \nicefrac{W}{8}$
spatial tokens. This corresponds to 4{,}800 tokens at the common COCO
benchmark resolution of $640{\times}480$ and over 32{,}000 at Full~HD
resolution.
Each token is processed by every cross-attention and a feed-forward
layer across all decoder stages, so activation traffic and decoder
latency grow linearly with image area.

CompactInst addresses the bottleneck with two complementary
modifications to the decoder pipeline.
First, we apply spatial max-pooling to reduce the pixel features from
$\nicefrac{1}{8}$ to $\nicefrac{1}{16}$ resolution before the
transformer decoder, cutting the token count by $4{\times}$.
Second, after decoding, a Refinement Head upsamples
the decoded features and fuses them with the original high-resolution
pixel-decoder features via a skip connection, recovering the
fine-grained spatial detail needed for accurate mask prediction.
Across six resolutions, the decoder-only latency gap between FastInst
and CompactInst on Jetson Orin~Nano widens with input size, peaking at
Cityscapes.
Under TensorRT FP16, CompactInst delivers up to a $4.74{\times}$
decoder and $2.62{\times}$ model speedup on Jetson Orin~Nano while
closely matching accuracy on COCO~\texttt{val2017} ($-0.4$\,AP) and
Cityscapes~\texttt{test} ($+0.2$\,AP).
A higher-resolution variant (M$\nicefrac{1}{4}$) with $2.57{\times}$
more decoder FLOPs still runs its decoder $1.74{\times}$ faster
than FastInst's on the most bandwidth-constrained device, further
confirming that activation traffic, not arithmetic cost, governs
decoder latency.

\pagebreak
\noindent Our contributions are summarized as follows:
\begin{enumerate}[topsep=0pt,itemsep=1pt,parsep=0pt,partopsep=0pt]
  \item We identify the transformer decoder of query-based segmentation
    models as memory-bandwidth-bound at deployment precision,
    established via an FP16 + tensor-core roofline analysis, and trace
    the root cause to cross-kernel activation traffic.
  \item We provide systematic TensorRT FP16 latency profiling across
    three GPUs spanning a
    $6.6{\times}$ DRAM bandwidth range (RTX A4000, Jetson AGX~Orin,
    Jetson Orin~Nano) and six input resolutions, showing larger
    token-reduction gains at higher resolutions and on lower-bandwidth
    devices.
  \item We propose CompactInst, a minimal architectural exploit
    ($4{\times}$ token-reducing pooling and a refinement
    head) that delivers a $4.74{\times}$ decoder-only and
    $2.62{\times}$ model speedup on Jetson Orin~Nano at
    Cityscapes resolution while closely matching accuracy.
  \item We show that the diagnosis generalises beyond FastInst: in a
    matched Mask2Former ablation, our design reduces decoder latency by
    $61\%$ despite
    adding $231\%$ decoder FLOPs, whereas PEM~\cite{PEM} has $32\%$
    fewer decoder FLOPs but increases decoder latency by $13\%$,
    demonstrating that
    FLOPs alone do not predict edge latency.
\end{enumerate}

\section{Related Work}
\label{sec:related}

\noindent\textbf{Instance segmentation.}
Instance segmentation requires jointly detecting objects and predicting
per-instance masks.
Two-stage methods such as Mask~R-CNN~\cite{MaskRCNN} first generate
region proposals and then refine bounding boxes and masks within each
region.
Single-stage approaches remove the proposal step:
SOLOv2~\cite{SOLOv2} predicts masks on a spatial grid via dynamic
convolutions,
CondInst~\cite{CondInst} generates instance-conditioned filters, and
SparseInst~\cite{SparseInst} activates a sparse set of instance-aware
features for real-time inference.
These convolutional methods achieve competitive latency and accuracy
but are inherently task-specific and do not naturally generalise to
panoptic or open-set segmentation.

\noindent\textbf{Query-based instance segmentation.}
DETR~\cite{DETR} introduced learnable object queries decoded by a
transformer, formulating detection as a direct set-prediction problem
and eliminating hand-crafted components such as non-maximum
suppression.
MaskFormer~\cite{MaskFormer} extended this paradigm to segmentation
by predicting a class label and mask embedding per query, enabling a
single architecture for semantic, instance, and panoptic segmentation.
Mask2Former~\cite{Mask2Former} incorporated multi-scale pixel features
at $\nicefrac{1}{32}$, $\nicefrac{1}{16}$, and $\nicefrac{1}{8}$
resolution into the decoder and introduced masked cross-attention.
FastInst~\cite{FastInst} builds on Mask2Former with
instance-activation-guided queries, a dual-path decoder that jointly
updates pixel features alongside queries, and ground-truth
mask-guided learning, achieving real-time inference with only three
decoder layers and a lightweight PPM-FPN pixel decoder.
Unlike Mask2Former, FastInst's pixel features reside at
$\nicefrac{1}{8}$ resolution and serve as both the decoder's tokens
and the final mask features, making their count a direct determinant
of decoder cost and mask quality.
FastInst's reported real-time numbers, like those of its query-based
predecessors, are measured on a high-bandwidth workstation GPU
(V100, 900\,GB/s) at FP32.
CompactInst directly modifies the FastInst pipeline, as detailed in
\cref{sec:method}.

\noindent\textbf{Token efficiency in vision transformers.}
Prior token-reduction methods for ViT classification
backbones~\cite{ToMe,EViT,DynamicViT,TokenLearner} rely on global class
tokens or uniform patch structure in backbone encoders.
At the pixel-decoder level,
HyPiDecoder~\cite{HyPiDecoder} replaces MSDeformAttn layers in the
Mask2Former pixel decoder with convolution-based FPN blocks, noting
that MSDeformAttn suffers from a low compute-to-memory access ratio.
FastInst~\cite{FastInst} already observed that swapping the
MSDeformAttn pixel decoder for a lightweight PPM-FPN yields latency
gains far exceeding the FLOP reduction, yet did not identify the root
cause.
Inside the transformer decoder itself, MOBIUS~\cite{MOBIUS} is the
closest scale-reduction prior, feeding a single stride-16 scale
to the decoder of large multi-modal generalist models.
PEM~\cite{PEM} reduces the pixel tokens consumed by cross-attention
through adaptive per-query prototype selection, making it the closest
in-attention token-reduction prior.
We compare PEM with the standard Mask2Former decoder and our design in
a matched single-path ablation that varies only the decoder
(\cref{sec:exp:ablations}).
SparseRefine~\cite{SparseRefine} shares the downsample-then-refine
principle, but applies it at the image level for semantic segmentation.
CompactInst instead applies parameter-free max-pooling within a
lightweight single-task pipeline, recovers detail through a refinement
head, and identifies the memory-bandwidth bottleneck that makes token
reduction translate into latency gains. None of these methods examine
this mechanism.

\noindent\textbf{Memory-bandwidth bottleneck of vision transformers.}
Ivanov~\etal~\cite{Ivanov2021DataMovement} quantified how transformer
workloads can be dominated by data movement rather than arithmetic.
In the vision domain, EfficientViT~\cite{Liu2023EfficientViT} and
LITv2~\cite{Pan2022LITv2HiLo} independently show that memory-bound
operations, rather than FLOPs, govern practical ViT latency, while
FLAT~\cite{Kao2023FLAT} formally characterises attention as severely
memory-bounded on hardware accelerators due to limited data reuse and
quadratic memory-footprint growth.
More recently, QUILL~\cite{QUILL} uses GPU roofline analysis to identify
low arithmetic intensity and irregular-memory-access bottlenecks in the
MSDeformAttn encoder and decoder of Deformable DETR, addressing them
with a custom fused accelerator.
The most prominent kernel-level response,
FlashAttention~\cite{Dao2022FlashAttention} fuses the attention core
($\mathrm{softmax}(QK^\top)V$) so that the attention matrix is not
materialised in off-chip memory, and is in principle applicable to the
cross-attention in our decoder.
This is complementary to our approach: attention fusion reduces data
movement within the attention core, whereas reducing the pixel-token
count also reduces traffic outside it, including traffic from the
pixel-path FFN, normalisation, and elementwise residual additions.
\Cref{sec:exp:traffic} quantifies the respective traffic scope of
attention fusion and pooling.
While prior work establishes memory-boundedness for ViT encoders,
self-attention, and deformable-attention object detectors, the analogous
diagnosis for the standard cross-attention transformer decoder of
query-based segmentation models has, to our knowledge, not been
performed on edge GPUs, whose DRAM bandwidth is substantially lower.
The mechanism is distinct from the well-known memory wall of
autoregressive LLM decoding, where each generated token requires large
model weights and the growing KV-cache to be read from memory.
Our segmentation decoder is non-autoregressive and has no persistent
KV-cache; its dominant measured traffic instead comes from
pixel-token-dependent intermediate activations materialised between
decoder kernels.
\Cref{sec:bottleneck} presents this diagnosis.

\section{The Memory-Bandwidth Bottleneck}
\label{sec:bottleneck}

The roofline model~\cite{Williams2009Roofline} relates achievable
throughput to \emph{arithmetic intensity} (AI), measured in FLOPs per
byte moved. A kernel must both execute its operations, bounded by the
peak compute rate $C_\text{peak}$, and move data to and from DRAM,
bounded by the peak memory bandwidth $B_\text{peak}$. Its achievable
performance is therefore
$\min(\mathrm{AI} \cdot B_\text{peak},\, C_\text{peak})$.
The two bounds meet at the \emph{ridge point}
$\mathrm{AI}^* = C_\text{peak} / B_\text{peak}$: a kernel with
$\mathrm{AI} < \mathrm{AI}^*$ performs too few operations per byte
fetched to keep the compute units busy. Its latency is governed by
memory traffic, so reducing FLOPs alone would not reduce runtime.
Because CUDA-only and tensor-core (TC) kernels have very different
peak rates, we follow
Yang~\etal~\cite{Yang2020HierarchicalRoofline} in using a separate
compute ceiling per execution unit: two ridges over a shared bandwidth
floor, with each kernel class compared against its own ridge.

Although our deployment targets are edge devices, we profile on an
RTX~A4000: it is the only platform in our study that
exposes DRAM traffic counters (\texttt{dram\_\_bytes.sum}),
whereas the Tegra iGPUs of both Jetsons expose only L2-level traffic,
an upper bound on DRAM bytes.
The roofline below therefore uses the A4000 measurements; edge behaviour
is tested separately in the deployment experiments.
With the A4000 configured at a $1.56$\,GHz SM clock and a
$7.0$\,GHz memory clock, the nominal FP16 ceilings are:
\begin{equation}
\label{eq:ceilings}
B_\text{peak} = 448\;\text{GB/s},\quad
\begin{cases}
  C_\text{CUDA} = 19.2\;\text{TFLOP/s} &\Rightarrow\;
    \mathrm{AI}^*_\text{CUDA} = 42.8\;\text{FLOPs/B} \\[2pt]
  C_\text{TC} = 76.7\;\text{TFLOP/s} &\Rightarrow\;
    \mathrm{AI}^*_\text{TC} = 171.1\;\text{FLOPs/B}
\end{cases}
\end{equation}

\paragraph{Methodology.}
We use NVIDIA Nsight Compute (\texttt{ncu}) to profile the transformer
decoder in isolation (three dual-path decoder layers; backbone, pixel
decoder, and post-processing excluded).
The model is run under pure FP16 (\texttt{model.half()}) so that the
ceilings above apply directly, and memory traffic is measured as true
DRAM bytes (\texttt{dram\_\_bytes.sum}).
The standard Mask2Former~\cite{Mask2Former} decoder is also profiled
under the same protocol to assess whether the diagnosis is specific to
FastInst.
FLOPs are counted from absolute hardware instruction counts, with
\texttt{tc16} and \texttt{tc8} denoting the observed $m16n8k16$ and
$m16n8k8$ tensor instructions:
\begin{equation}
\label{eq:flops}
\text{FLOPs} = \underbrace{\texttt{fadd+fmul+2ffma}}_{\text{FP32}}
             + \underbrace{\texttt{hadd+hmul+2hfma}}_{\text{FP16}}
             + \underbrace{\texttt{4096tc16+2048tc8}}_{\text{TC}} .
\end{equation}
Results are grouped into three aggregates per model:
(1)~\emph{Combined} (all kernels),
(2)~\emph{CUDA-only} (kernels with $\texttt{tc}=0$), and
(3)~\emph{TC-only} (kernels with $\texttt{tc}>0$).
Each aggregate's AI is total FLOPs over total bytes within its kernel
set; comparing CUDA-only against $\mathrm{AI}^*_\text{CUDA}$ and
TC-only against $\mathrm{AI}^*_\text{TC}$ avoids conflating two
execution domains.

\begin{figure}[t]
  \centering
  \subfigure[FP16 (tensor cores active)]{%
    \includegraphics[width=0.49\linewidth]{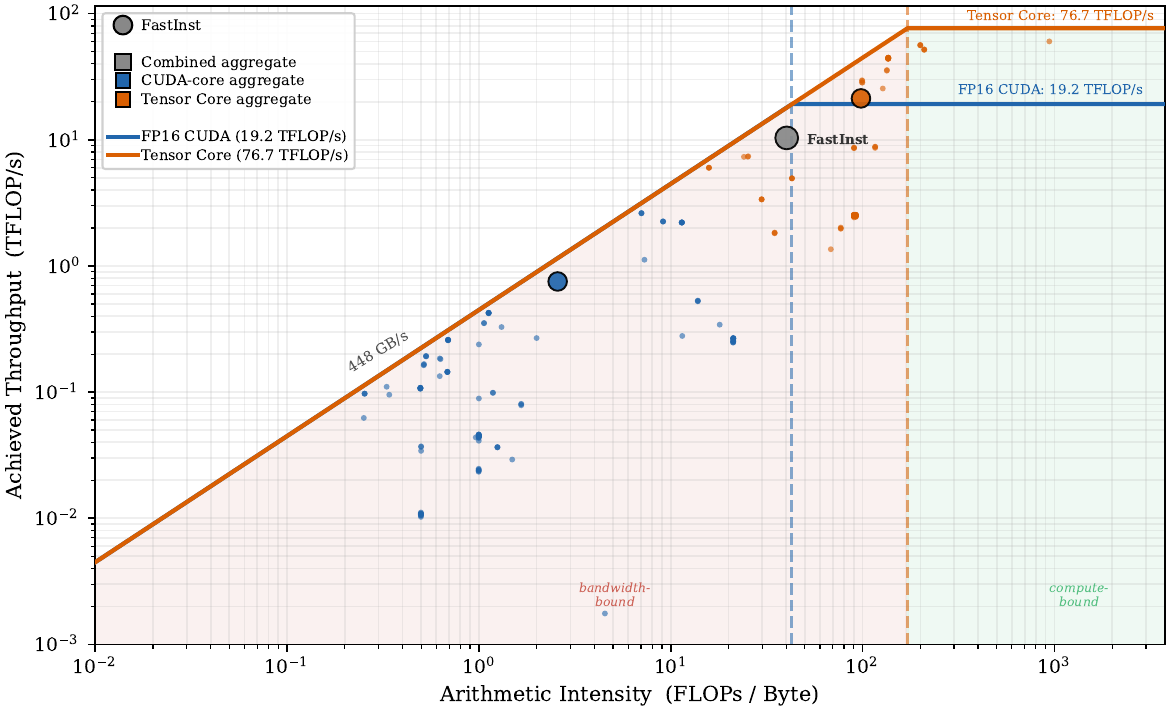}%
    \label{fig:roofline-fp16}%
  }\hfill
  \subfigure[FP32 (tensor cores idle)]{%
    \includegraphics[width=0.49\linewidth]{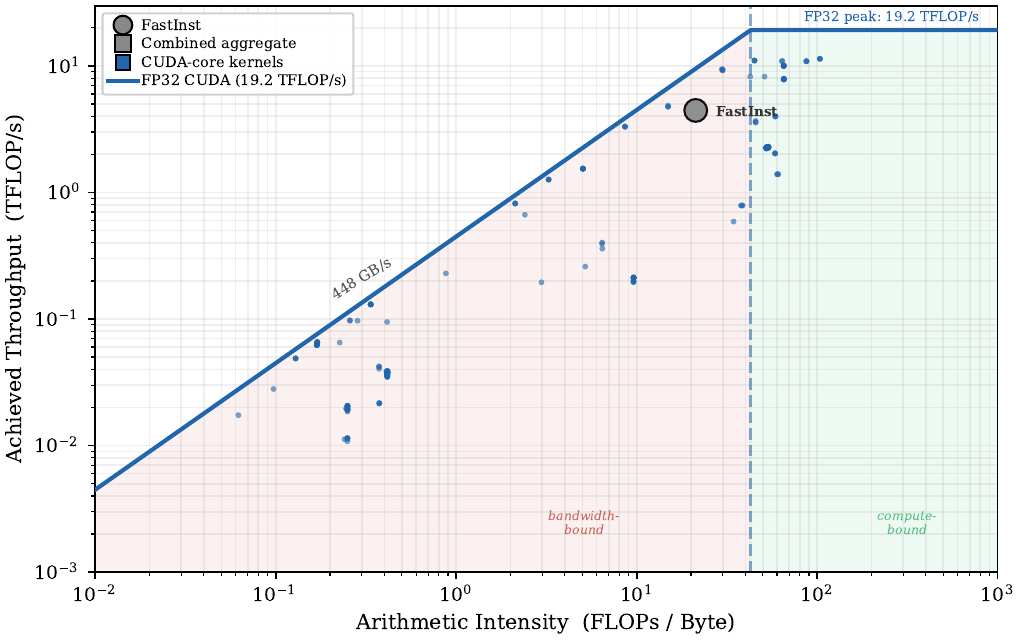}%
    \label{fig:roofline-fp32}%
  }
  \caption{%
    DRAM rooflines of the FastInst decoder on the
    RTX~A4000 ($1088{\times}1920$; bytes =
    \texttt{dram\_\_bytes.sum}).
    Large markers denote aggregate values, and both panels use a
    $448$ GB/s bandwidth slope.
    (a) Under FP16, the CUDA-only activation aggregate (blue)
    and tensor-core aggregate (orange) lie below their respective ridges
    ($42.8$ and $171.1$ FLOPs/B), as do most individual kernels.
    (b) Under FP32, the single ridge is $42.8$ FLOPs/B and
    the GEMM cluster to its right is compute-bound.%
  }
  \label{fig:roofline}
\end{figure}

\paragraph{FP16 roofline results.}
At Full HD ($1088{\times}1920$),
\cref{fig:roofline-fp16,tab:dram-aggregates} show that both
decoders operate deeply in the bandwidth-bound regime under deployment
precision. Decomposing by execution unit reveals where the traffic
originates. The CUDA-only aggregates, including elementwise,
LayerNorm, and softmax kernels, are severely bandwidth-bound at
$2.5\%$ to $6\%$ of the CUDA ridge, yet these kernels account for
$53\%$ to $70\%$ of decoder time and $61\%$ to $71\%$ of DRAM bytes.
They predominantly stream pixel-token activations between GEMMs, so
their traffic is overwhelmingly activation traffic. The largest
individual kernels sustain up to $90\%$ of nominal peak DRAM bandwidth,
effectively saturating the memory bus.
The GEMM-carrying TC kernels, by contrast, sit at $41\%$ to $57\%$ of
their ridge (moderately bound) and move the minority of the bytes;
kernels that are individually compute-bound account for only $9.3\%$
of FastInst's decoder time.
The bottleneck is therefore primarily cross-kernel activation traffic,
rather than low GEMM intensity.

\begin{table}[t]
  \centering
  \caption{%
    DRAM roofline aggregates on the RTX~A4000 (FP16, $1088{\times}1920$): 
    arithmetic intensity, kernel time, and DRAM bytes with within-decoder
    shares. Per-path rows show \% of the corresponding ridge (CUDA-only
    vs.\ $42.8$; TC vs.\ $171.1$); Combined AI is left unnormalised.%
  }
  \label{tab:dram-aggregates}
  \footnotesize
  \renewcommand{\arraystretch}{1.15}
  \begin{tabular}{llccc}
    \hline
    Model & Aggregate & AI [FLOPs/B] (\%\,ridge) & Time [ms] (\%) & DRAM [MB] (\%)\\
    \hline
    \multirow{3}{*}{FastInst~\cite{FastInst}}
      & CUDA-only & 2.6 (6.0\%) & 11.9 (53\%) & 3474 (61\%)\\
      & TC        & 98 (57\%)   & 10.4 (47\%) & 2262 (39\%)\\
      & Combined  & 40           & 22.3        & 5736\\
    \hline
    \multirow{3}{*}{Mask2Former~\cite{Mask2Former}}
      & CUDA-only & 1.1 (2.5\%) & 15.8 (70\%) & 4938 (71\%)\\
      & TC        & 70 (41\%)   & 6.8 (30\%)  & 1987 (29\%)\\
      & Combined  & 21           & 22.7        & 6925\\
    \hline
  \end{tabular}
\end{table}

\paragraph{FP32 roofline results.}
Under FP32 (tensor cores idle) the A4000 peak is $19.2$\,TFLOP/s with
a single ridge at $42.8$\,FLOPs/B. The decoder operates in a mixed regime:
kernels individually above the ridge account for $55.7\%$ of decoder
time. The GEMM-class kernels collectively reach AI $52.7$ ($123\%$ of
the ridge), with the largest individual GEMM reaching AI $103.6$
($242\%$), whereas the non-GEMM kernels have aggregate AI $1.77$ and
account for $36.9\%$ of decoder time. The combined AI of $21.2$
therefore obscures these two distinct kernel regimes
(\cref{fig:roofline-fp32}).

Under FP16, the tensor-core ceiling is $4{\times}$ higher while memory
bandwidth is unchanged. The share of decoder time in kernels above
their corresponding ridge falls to $9.3\%$, while the non-GEMM share
rises to $53.3\%$. FP16 tensor acceleration therefore does not create
the low-intensity traffic; it makes the existing cross-kernel
activation traffic the dominant runtime cost.

Reducing the pixel-token count is the architectural lever that shrinks
cross-kernel activation traffic; the design in \cref{sec:method}
exploits exactly this.

\section{CompactInst: a Minimal Bottleneck Exploit}
\label{sec:method}

CompactInst directly targets the cross-kernel activation traffic
identified in \cref{sec:bottleneck} through a minimal architectural
change to FastInst~\cite{FastInst}, without otherwise altering the
pipeline.

\begin{figure*}[t]
  \centering
  \includegraphics[width=\textwidth]{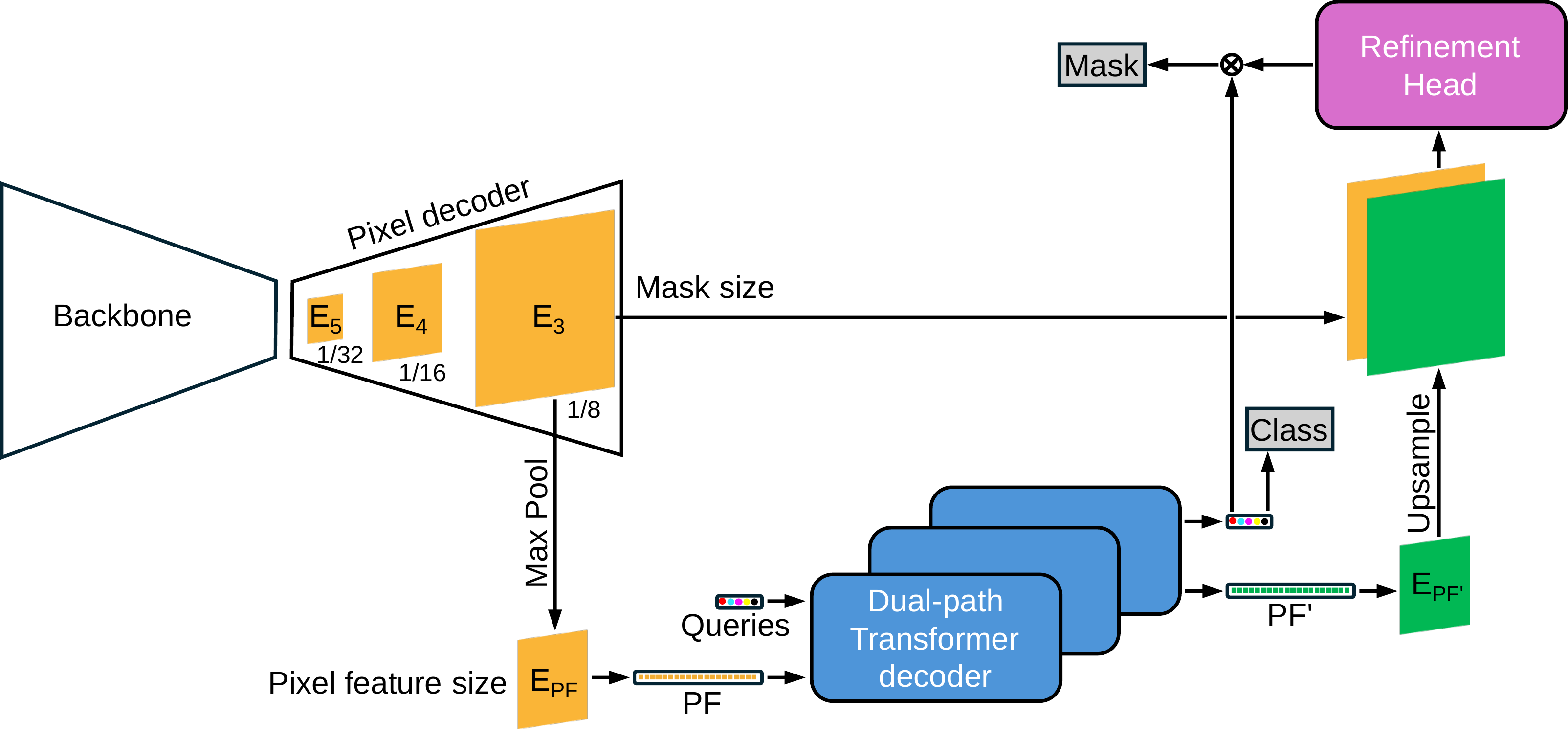}
  \caption{%
    CompactInst architecture.
    Building on FastInst~\cite{FastInst}, CompactInst consists of a
    backbone, a PPM-FPN pixel decoder~\cite{SparseInst}, and a
    dual-path transformer decoder.
    The pixel decoder produces $\mathbf{E}_3$ at $\nicefrac{1}{8}$
    resolution; adaptive max-pooling reduces it to
    $\mathbf{E}_\text{PF}$ at $\nicefrac{1}{16}$, which is flattened
    into pixel-feature tokens~$\text{PF}$ for the $D{=}3$ decoder
    layers.
    After the last layer, the refined features
    $\mathbf{E}_{\text{PF}'}$ are upsampled back to $\nicefrac{1}{8}$
    and concatenated with $\mathbf{E}_3$; two convolutional layers
    (Refinement Head) then produce the final mask features.
    This allows the decoder to operate on $4{\times}$ fewer pixel
    tokens while recovering spatial detail through the
    high-resolution skip connection.
  }
  \label{fig:arch}
\end{figure*}

As described in \cref{sec:related}, CompactInst retains FastInst's
ResNet backbone, PPM-FPN pixel decoder, instance-activation-guided
and auxiliary queries, positional embeddings, and three dual-path
transformer decoder layers (\cref{fig:arch}); only the pixel-feature
path immediately before and after the transformer decoder changes.

\pagebreak
\noindent\textbf{Pixel-token reduction.}
CompactInst applies parameter-free adaptive max-pooling to the
pixel-decoder features before the transformer decoder, producing
$\mathbf{E}_\text{PF}$ at target pixel-feature stride $s_\text{PF}$
(e.g.\ $16$ or $32$), and flattens the result into the pixel-token
sequence consumed by the subsequent decoder layers:
\begin{equation}
  \mathbf{E}_\text{PF}
    = \operatorname{AdaptiveMaxPool2d}\!\bigl(
        \mathbf{E}_3,\;
        \bigl(\tfrac{H}{s_\text{PF}},\,\tfrac{W}{s_\text{PF}}\bigr)
      \bigr),
  \label{eq:pool}
\end{equation}
where $s_\text{PF}$ is the target pixel-feature stride.
The default setting is $s_\text{PF}{=}16$, which reduces the token
count by $4{\times}$; setting $s_\text{PF}{=}32$ yields a $16{\times}$
reduction.
We abbreviate mask resolution as~M and pixel-feature stride as~PF;
the default configuration is thus
M\,$\nicefrac{1}{8}$\,PF\,$\nicefrac{1}{16}$.
Pooling-operator and stride choices are evaluated in
\cref{sec:exp:ablations}.

\noindent\textbf{Refinement Head.}
Rather than predicting masks directly from the final reduced-resolution
token sequence $\text{PF'}$, CompactInst applies a Refinement Head only
after the last decoder layer, leaving all intermediate layers at the
reduced pixel-feature resolution. The head unflattens $\text{PF'}$ into
$\mathbf{E}_{\text{PF}'}$, upsamples it to the mask resolution, and
concatenates it channel-wise with the original high-resolution
pixel-decoder features ($\mathbf{E}_3$ at $\nicefrac{1}{8}$, or
$\mathbf{E}_2$ at $\nicefrac{1}{4}$ for M$\nicefrac{1}{4}$).
Two $3{\times}3$ Conv-GN-ReLU layers produce the final mask features:
the first projects the $2C$ concatenated channels to $C$, and the
second preserves that dimensionality. The head replaces FastInst's
final-layer linear projection and adds ${\sim}1.7$\,M parameters
(${\sim}5\,\%$ of 35.9\,M); its two convolutions add mainly FLOPs
rather than repeated pixel-token traffic.

Note that in architectures without a dual-path decoder (i.e.\ where
pixel features are not co-optimised with queries), one could omit the
Refinement Head and predict masks from separate higher-resolution
pixel-decoder features; Mask2Former~\cite{Mask2Former} evaluated this
setting ($\nicefrac{1}{16}$ decoder, $\nicefrac{1}{4}$ masks) and
observed ${\sim}1$\,AP drop (Table~4d in~\cite{Mask2Former}).

\section{Experiments}
\label{sec:experiments}

\subsection{Setup}
\label{sec:exp:setup}

\paragraph{Datasets and metrics.}
We train on COCO~\cite{COCO} \texttt{train2017} and
Cityscapes~\cite{Cordts2016Cityscapes} \texttt{train} splits, and
evaluate on COCO \texttt{val2017} and Cityscapes \texttt{val} and
\texttt{test} using standard mask AP metrics. For Cityscapes, we
additionally apply the COCO evaluation protocol on \texttt{val} to
obtain size-stratified AP.

\paragraph{Implementation details.}
CompactInst extends FastInst~\cite{FastInst} in
Detectron2~\cite{Detectron2}.
All models use a ResNet-50~\cite{ResNet} backbone pre-trained on
ImageNet~\cite{ImageNet} and the PPM-FPN pixel
decoder~\cite{FastInst}.
We adopt FastInst's training recipe (AdamW~\cite{AdamW}, bipartite
matching loss, AMP) with the following hardware adaptations.
For COCO, we train with batch size~16 on $4{\times}$~NVIDIA T4 GPUs,
matching FastInst's original setting.
For Cityscapes, we use a single RTX~3090 with batch size~2;
the learning rate is scaled linearly with the batch size and the
schedule steps, warmup, and total iterations are scaled
inversely~\cite{Goyal2017LinearScaling}.

\paragraph{Hardware platforms.}
Latency is measured on three platforms spanning a $6.6{\times}$ range
in memory bandwidth:
\begin{itemize}[nosep]
  \item \textbf{RTX A4000}: 448\,GB/s GDDR6 (workstation GPU);
  \item \textbf{Jetson AGX Orin}: 204.8\,GB/s LPDDR5 (high-end edge);
  \item \textbf{Jetson Orin Nano}: 68\,GB/s LPDDR5 (low-power edge).
\end{itemize}

\paragraph{Deployment protocol.}
All latency numbers are obtained from TensorRT FP16 static engines
with CUDA-event timing, 50 warmup and 200 timed iterations
at batch size~1.
This replaces the FP32 measurements obtained using standard PyTorch/
Detectron2 inference used in earlier work and matches the deployment
precision analysed in \cref{sec:bottleneck}.
Engines are built from ONNX using TensorRT's FP16 builder flag and no
project-specific plugins; additional details are provided in the
supplementary.
Latency is reported at two measurement scopes:
(1)~\emph{decoder-only} timings isolate the transformer decoder from
backbone, pixel decoder, and post-processing, enabling direct
comparison of the component targeted by our intervention;
(2)~\emph{model} timings (ResNet-18 backbone, all network components,
no post-processing) capture the full-network benefit visible to a
practitioner; we use ResNet-18 here because the ResNet-50 models
exceed Orin~Nano's available memory during TensorRT inference at higher
resolutions.
Post-processing is excluded because it runs outside the compiled
engine and is unchanged between FastInst and CompactInst. It is a
separate, untargeted bandwidth bottleneck outside the decoder:
upsampling, thresholding, and mask scoring move full-resolution mask
tensors with little arithmetic. In our AGX~Orin AMP breakdown at
Cityscapes resolution, post-processing is the second-largest component
of FastInst and the largest component of CompactInst for both R-18 and
R-50 (supplementary).

\subsection{Decoder-only speedup}
\label{sec:exp:decspeedup}

The roofline analysis (\cref{sec:bottleneck}) identifies activation
traffic as the bottleneck. We therefore examine how the benefit of
token reduction changes with device bandwidth.
We test this by compiling the FastInst and CompactInst decoders
(\cref{sec:method}) as TensorRT FP16 static engines and timing them
in isolation across six resolutions on each platform.

\begin{figure}[t]
  \centering
  \includegraphics[width=\linewidth]{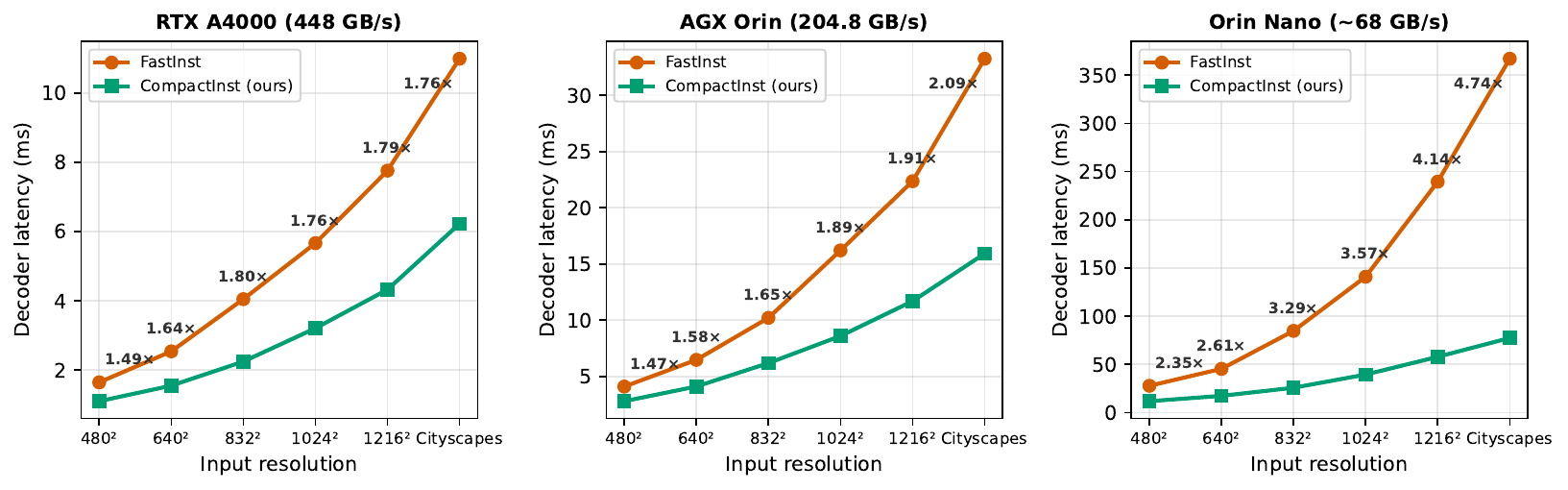}
  \caption{%
    Decoder-only TensorRT FP16 latency across three GPUs spanning
    $6.6{\times}$ DRAM bandwidth. Annotations show FastInst/CompactInst
    latency. The speedup generally increases as device bandwidth decreases
    and peaks at $4.74{\times}$ on Orin~Nano, supporting the bandwidth-bound
    diagnosis in \cref{sec:bottleneck}.%
  }
  \label{fig:trt-3gpu}
\end{figure}

\Cref{fig:trt-3gpu} reports decoder-only latency with annotated speedups.
CompactInst achieves speedups from $1.49{\times}$ to $1.80{\times}$
on the A4000 (448\,GB/s), from $1.47{\times}$ to $2.09{\times}$ on
AGX~Orin (204.8\,GB/s), and from $2.35{\times}$ to $4.74{\times}$ on
Orin~Nano (68\,GB/s).
The speedup generally increases as device bandwidth decreases, with the
largest gains on Orin~Nano. This cross-device trend is consistent with
the activation-traffic mechanism, which \cref{sec:exp:traffic} supports
directly with measured DRAM traffic.

Within each device the speedup also grows with resolution, most
steeply on Orin~Nano ($2.35{\times}{\to}4.74{\times}$).
A likely contributor is its halved L2 (2\,MB vs.\ 4\,MB on A4000 and
AGX~Orin), causing FastInst's larger working set to spill to DRAM
sooner.

\subsection{Model-level deployment latency}
\label{sec:exp:e2e}

\begin{figure}[!h]
  \centering
  \includegraphics[width=\linewidth]{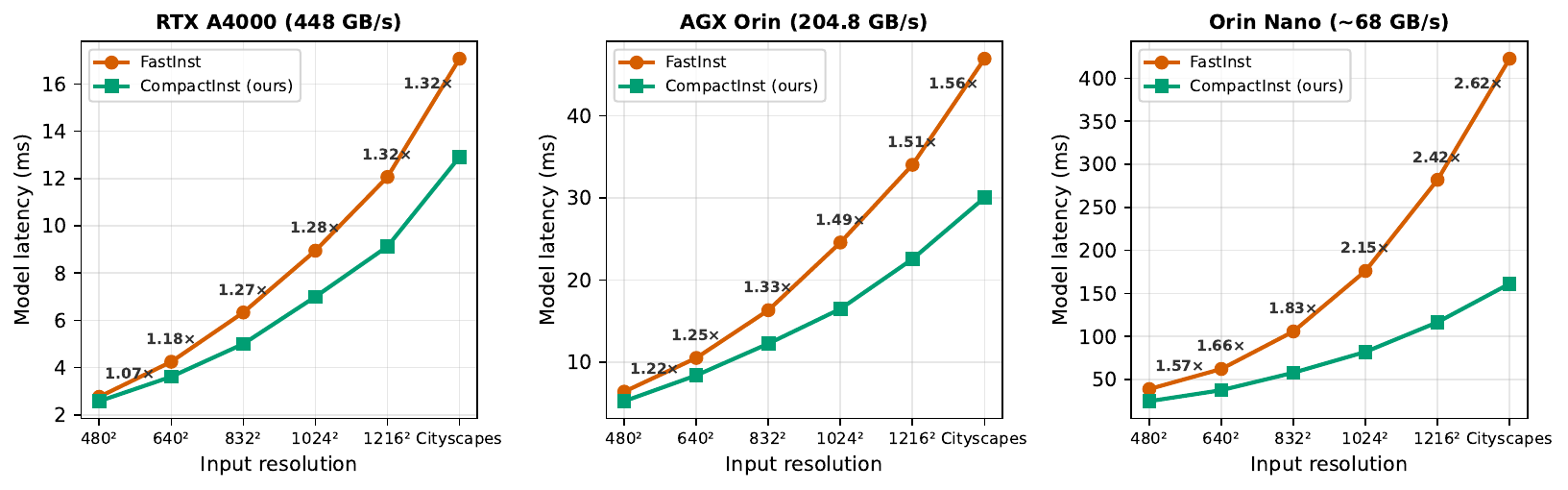}
  \caption{%
    Model-level TensorRT FP16 latency (no post-processing) across
    three GPUs. Annotations show FastInst/CompactInst latency. Speedup
    increases with resolution and is largest on Orin~Nano, the
    lowest-bandwidth device.%
  }
  \label{fig:e2e-3gpu}
\end{figure}

\Cref{fig:e2e-3gpu} reports model latency and annotated speedups
(ResNet-18 backbone, no post-processing) across all three platforms.
On Orin~Nano, CompactInst delivers up to $2.62{\times}$ at Cityscapes
resolution. Across the measured platforms, gains generally increase with
input resolution and are largest on Orin~Nano
(from $1.07{\times}$ to $1.32{\times}$ on A4000 and from
$1.22{\times}$ to $1.56{\times}$ on AGX~Orin).
CompactInst (default) saves only $7\,\%$ FLOPs (190.8\,G vs.\
205.8\,G) yet achieves up to $2.62{\times}$ speedup. The latency gain
therefore far exceeds what arithmetic savings alone would predict.

\subsection{More FLOPs, less time: the M$\nicefrac{1}{4}$ counterfactual}
\label{sec:exp:m14}

The M$\nicefrac{1}{4}$ variant keeps the $\nicefrac{1}{16}$ pooled
decoder path but upsamples mask features to $\nicefrac{1}{4}$, adding a
heavier refinement head (part of the transformer decoder; included in
all decoder FLOP and latency numbers reported in this paper).
Its decoder FLOPs are $2.57{\times}$ those of FastInst at Cityscapes
resolution (264.9\,G vs.\ 103.2\,G).

At Cityscapes resolution on Orin~Nano, M$\nicefrac{1}{4}$'s decoder
runs $1.74{\times}$ faster than FastInst's despite
$2.57{\times}$ more decoder FLOPs. At the same resolution on the A4000
($6.6{\times}$ more bandwidth), the corresponding
FastInst/M$\nicefrac{1}{4}$ ratio is $0.75{\times}$, i.e.\ the variant is
slower.
At the model level, M$\nicefrac{1}{4}$ first exceeds parity on
Orin~Nano at the measured $1024^{2}$ resolution and reaches
$1.33{\times}$ at Cityscapes resolution, despite having
$2.17{\times}$ as many total FLOPs as FastInst (446.5\,G vs.\
205.8\,G). It remains slower at every measured resolution on the A4000
and AGX~Orin. Full curves are reported in the supplementary.
This device-dependent result shows that FLOP count alone is
insufficient to predict latency and is consistent with
memory traffic becoming the binding constraint as device bandwidth
decreases. Together with the directly measured DRAM traffic in
\cref{sec:exp:traffic}, it shows that bytes moved, rather than
operations performed, govern latency on bandwidth-limited hardware.

\subsection{Where the speedup comes from: measured DRAM traffic}
\label{sec:exp:traffic}

The latency evidence above is indirect. We profile CompactInst's
decoder under the same DRAM-measurement protocol as \cref{sec:bottleneck} to
measure the traffic directly (\cref{tab:ci-traffic}).
Pooling removes $64\%$ of all decoder DRAM bytes
($5.74 \to 2.09$\,GB, $2.74{\times}$ fewer than FastInst in
\cref{tab:dram-aggregates}).
Most of this reduction occurs in the CUDA-only activation bucket,
which shrinks from $3474$ to $1275$\,MB while its aggregate AI remains
similarly low ($2.6 \to 2.3$).
Pooling reduces pixel-token activation volume by construction; the
A4000 traffic measurement and the widening speedup on lower-bandwidth
GPUs (\cref{fig:trt-3gpu,fig:e2e-3gpu}) are consistent with this
mechanism.

\begin{table}[h]
  \centering
  \caption{%
    CompactInst decoder aggregates under the same DRAM-measurement
    protocol as \cref{tab:dram-aggregates} (RTX~A4000, FP16, Full HD
    $1088{\times}1920$).
    Relative to FastInst, total DRAM bytes drop $2.74{\times}$.
    Combined AI is reported without normalisation to a single ridge.%
  }
  \label{tab:ci-traffic}
  \footnotesize
  \renewcommand{\arraystretch}{1.15}
  \begin{tabular}{llccc}
    \hline
    Model & Aggregate & AI [FLOPs/B] (\%\,ridge) & Time [ms] (\%) & DRAM [MB] (\%)\\
    \hline
    \multirow{3}{*}{CompactInst (ours)}
      & CUDA-only & 2.3 (5.3\%) & 5.8 (53\%)  & 1275 (61\%)\\
      & TC        & 220 (129\%) & 5.1 (47\%)  & 819 (39\%)\\
      & Combined  & 87           & 11.0        & 2094\\
    \hline
  \end{tabular}
\end{table}

\paragraph{How much can kernel fusion reach?}
The same measurements answer the question raised in
\cref{sec:related}: at most $29\%$ of FastInst's decoder DRAM bytes
($1.67$ of $5.74$\,GB) lie in the attention chain potentially reachable
by FlashAttention-style fusion. The remaining $71\%$ lies outside the
attention core, in operations such as the pixel-path FFN, normalisation,
and elementwise residual additions.
By reducing the token count upstream, pooling removes $64\%$ of all
bytes; after pooling, only $19\%$ of CompactInst's decoder bytes lie in
this potentially fusible attention chain.

\subsection{Mask accuracy}
\label{sec:exp:accuracy}

\paragraph{COCO results.}
\Cref{tab:coco} compares methods on COCO \texttt{val2017} (R-50).
The CNN-based baselines (CondInst, SOLOv2, SparseInst) reach
$34.3$ to $36.0$\,AP, below the query-based methods in the table.
CompactInst (default) reaches $37.3$\,AP ($-0.4$ vs.\ FastInst),
with the gap concentrated in AP$_S$ ($16.3 \to 15.2$): on COCO,
pooling measurably costs small-object accuracy, a relevant
trade-off for edge perception, where small objects are often
safety-critical.
The PF\,$\nicefrac{1}{32}$ variant trades $1.4$\,AP for the fewest
decoder tokens.
M$\nicefrac{1}{4}$ achieves the highest AP in the table ($38.2$,
$+0.5$ over FastInst) and matches FastInst's AP$_S$ at $16.4$
($+0.1$).

\begin{table}[!htbp]
\centering
\caption{Instance segmentation on COCO \texttt{val2017}.
  All models use a ResNet-50 backbone, evaluated at $640{\times}853$.}
\label{tab:coco}
\setlength{\tabcolsep}{2.5pt}
\footnotesize
\renewcommand{\arraystretch}{1.15}
\begin{tabular}{@{}l cc cccccc@{}}
\hline
Method
  & Params
  & GFLOPs
  & AP & AP$_{50}$ & AP$_{75}$ & AP$_S$ & AP$_M$ & AP$_L$ \\
\hline
CondInst~\cite{CondInst}
  & 34.1\,M & 105.8
  & 35.8 & 56.2 & 37.8 & 14.3 & 39.0 & 55.7 \\
SOLOv2~\cite{SOLOv2}
  & 46.6\,M & 164.4
  & 36.0 & 56.0 & 38.4 & 12.0 & 39.2 & 58.3 \\
SparseInst~\cite{SparseInst}
  & 32.8\,M & 91.1
  & 34.3 & 55.2 & 36.1 & 14.2 & 36.8 & 51.9 \\
FastInst~\cite{FastInst}
  & 34.2\,M & 75.7
  & \underline{37.7} & \underline{59.2} & \underline{39.5} & \underline{16.3} & \underline{40.5} & 59.1 \\
\hline
CompactInst (ours)
  & 35.9\,M & 72.0
  & 37.3 & 58.4 & 39.0 & 15.2 & 40.1 & \textbf{60.0} \\
\quad w/ PF\,$\nicefrac{1}{32}$
  & 35.9\,M & 67.7
  & 36.3 & 57.4 & 38.1 & 14.0 & 39.1 & 59.2 \\
\quad w/ M\,$\nicefrac{1}{4}$\,PF\,$\nicefrac{1}{16}$
  & 36.5\,M & 136.4
  & \textbf{38.2} & \textbf{59.7} & \textbf{40.1}
  & \textbf{16.4} & \textbf{40.9} & \underline{59.9} \\
\hline
\end{tabular}
\end{table}

\paragraph{Cityscapes results.}
At native $1024{\times}2048$ resolution (\cref{tab:cityscapes}),
CompactInst matches FastInst on \texttt{test} ($+0.2$\,AP) and
\texttt{val} ($+0.6$\,AP).
The COCO small-object drop is the reason we additionally evaluate
Cityscapes \texttt{val} under the COCO protocol (starred columns),
which provides the size-stratified AP the native Cityscapes metric
lacks: it tests whether the AP$_S$ cost transfers to high-resolution
data.
It does not. AP$_S^{\star}$ improves by $+0.4$ ($+7\%$
relative): at native high resolution the max-pooled
$\nicefrac{1}{16}$ tokens still carry sufficient spatial detail for
small instances.
The small-object trade-off is thus dataset- and resolution-dependent,
and practitioners for whom small objects are critical should validate
it on their target domain.
Where it matters, M$\nicefrac{1}{4}$ is the mitigation: it lifts
AP$_S^{\star}$ to $7.7$ ($+2.0$ over FastInst, $+35\%$ relative) and
matches FastInst's AP$_S$ on COCO, while still running faster on
bandwidth-limited hardware (\cref{sec:exp:m14}).

\begin{table}[!htbp]
\centering
\caption{Instance segmentation on Cityscapes ($1024{\times}2048$).
  The starred ($^{\star}$) columns use the COCO evaluation protocol on
  \texttt{val} to provide size-stratified AP.}
\label{tab:cityscapes}
\setlength{\tabcolsep}{2.2pt}
\scriptsize
\renewcommand{\arraystretch}{1.15}
\begin{tabular}{@{}l c cccc cccc@{}}
\hline
 & & \multicolumn{2}{c}{\texttt{test}} &
   \multicolumn{2}{c}{\texttt{val}} &
   \multicolumn{4}{c}{\texttt{val}$^{\star}$ (COCO protocol)} \\
\cline{3-4} \cline{5-6} \cline{7-10}
Method & GFLOPs &
  AP & AP$_{50}$ & AP & AP$_{50}$ &
  AP$_{75}$ & AP$_S$ & AP$_M$ & AP$_L$ \\
\hline
FastInst~\cite{FastInst}
  & 308.1
  & 26.4 & \underline{48.5} & 30.0 & \underline{54.5}
  & 27.6 & 5.7 & \underline{25.0} & \underline{53.7} \\
\hline
CompactInst (ours)
  & 293.0
  & \underline{26.6} & \textbf{49.4} & \underline{30.6} & 53.4
  & \underline{28.5} & \underline{6.1} & \underline{25.0} & \textbf{56.7} \\
\quad w/ PF\,$\nicefrac{1}{32}$
  & 275.2
  & 25.6 & 47.1 & 28.2 & 51.6
  & 26.0 & 5.1 & 24.8 & 52.4 \\
\quad w/ M\,$\nicefrac{1}{4}$\,PF\,$\nicefrac{1}{16}$
  & 556.2
  & \textbf{27.6} & \textbf{49.4} & \textbf{31.2} & \textbf{56.4}
  & \textbf{30.1} & \textbf{7.7} & \textbf{27.0} & \textbf{56.7} \\
\hline
\end{tabular}
\end{table}

\subsection{Ablations}
\label{sec:exp:ablations}

\paragraph{Pooling strategy.}
We evaluate three strategies for reducing the pixel-decoder features
before the transformer decoder (\cref{tab:ablation_pool}).
\emph{Max-pooling} selects the strongest activation within each
spatial window independently per channel, preserving discriminative
peaks. The resulting feature vectors can be spatially inconsistent
because different channels may originate from different positions
within the window.
\emph{Average-pooling} computes the spatial mean per channel,
yielding smoother but potentially less discriminative features.
\emph{Foreground-guided (FG) selection} addresses the spatial
inconsistency of max-pooling: an auxiliary semantic-segmentation head
attached after the pixel decoder predicts a per-pixel foreground
probability (all instance classes vs.\ background); the feature map
is then divided into a grid matching the target resolution, and
within each cell the pixel with the highest FG probability is
selected, giving a spatially coherent feature vector across channels.

\begin{table}[!htbp]
\centering
\caption{Pooling strategy ablation on COCO \texttt{val2017}.
  All rows use CompactInst with the refinement head.}
\label{tab:ablation_pool}
\setlength{\tabcolsep}{3pt}
\footnotesize
\renewcommand{\arraystretch}{1.15}
\begin{tabular}{@{}ll cccccc@{}}
\hline
Strategy & Config &
  AP & AP$_{50}$ & AP$_{75}$ & AP$_S$ & AP$_M$ & AP$_L$ \\
\hline
Max-pool
  & M\,$\nicefrac{1}{8}$\,PF\,$\nicefrac{1}{16}$
  & \textbf{37.3} & \textbf{58.4} & \textbf{39.0} & \textbf{15.2} & \textbf{40.1} & \textbf{60.0} \\
Avg-pool
  & M\,$\nicefrac{1}{8}$\,PF\,$\nicefrac{1}{16}$
  & 37.1 & 58.2 & 38.7 & 15.0 & 40.0 & 59.3 \\
FG-guided
  & M\,$\nicefrac{1}{8}$\,PF\,$\nicefrac{1}{16}$
  & 36.9 & 58.2 & 38.3 & 14.9 & 39.9 & 59.3 \\
\hline
Max-pool
  & M\,$\nicefrac{1}{4}$\,PF\,$\nicefrac{1}{32}$
  & \textbf{37.1} & \textbf{58.1} & \textbf{39.1} & \textbf{15.6} & \textbf{39.7} & 58.8 \\
Avg-pool
  & M\,$\nicefrac{1}{4}$\,PF\,$\nicefrac{1}{32}$
  & 36.6 & 57.4 & 38.5 & 15.3 & 39.4 & \textbf{59.3} \\
\hline
\end{tabular}
\end{table}

Max-pooling consistently achieves the highest AP; the gap widens at
the more aggressive PF\,$\nicefrac{1}{32}$ ($+0.5$ over average),
suggesting that preserving peak activations becomes more important as
spatial reduction increases.
Foreground-guided selection does not improve over either pooling
baseline despite enforcing spatial consistency, indicating that
(1)~the per-channel max already preserves sufficiently discriminative
features and (2)~the refinement head recovers spatial coherence lost
during pooling by fusing with the original high-resolution FPN
features.

\paragraph{Refinement head.}
\Cref{tab:ablation_refine} isolates the refinement head's
contribution.
Removing it costs $-2.1$\,AP, with the largest drops in AP$_{75}$
($-2.7$) and AP$_S$ ($-2.9$), confirming that the skip connection
from the full-resolution FPN features is essential for recovering
fine mask detail lost during pooling.

\begin{table}[h]
\centering
\caption{Effect of the refinement head
  (M\,$\nicefrac{1}{8}$\,PF\,$\nicefrac{1}{16}$, COCO
  \texttt{val2017}).}
\label{tab:ablation_refine}
\setlength{\tabcolsep}{3pt}
\footnotesize
\renewcommand{\arraystretch}{1.15}
\begin{tabular}{@{}l cccccc@{}}
\hline
 & AP & AP$_{50}$ & AP$_{75}$ & AP$_S$ & AP$_M$ & AP$_L$ \\
\hline
w/o refinement & 35.2 & 56.6 & 36.3 & 12.3 & 38.1 & 59.6 \\
w/\ refinement & \textbf{37.3} & \textbf{58.4} & \textbf{39.0}
  & \textbf{15.2} & \textbf{40.1} & \textbf{60.0} \\
\hline
\end{tabular}
\end{table}

\begin{figure}[t]
  \centering
  \includegraphics[width=0.70\linewidth]{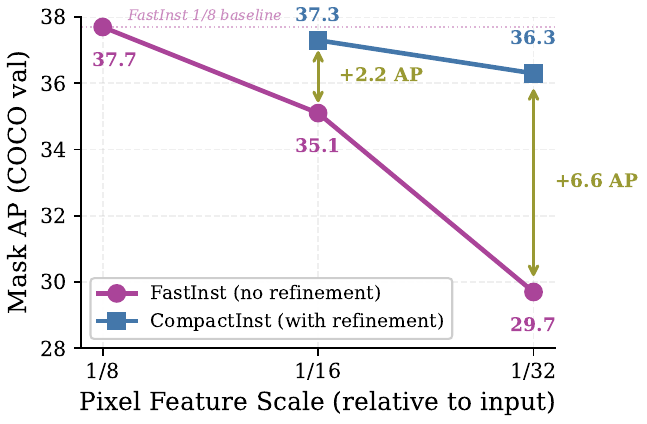}
  \caption{Mask AP vs.\ pixel-feature scale on COCO \texttt{val2017}.
    CompactInst (pool + refine) consistently outperforms FastInst's
    direct low-resolution decoding at every scale.}
  \label{fig:token_ablation}
\end{figure}

\paragraph{Pixel-feature resolution.}
CompactInst always upsamples the pixel decoder to $\nicefrac{1}{8}$
and max-pools to the target scale ($\nicefrac{1}{16}$ or
$\nicefrac{1}{32}$) for the transformer; the refinement head then
upsamples the refined low-resolution output back to $\nicefrac{1}{8}$,
concatenates it with the original $\nicefrac{1}{8}$ pixel-decoder
features, and applies two $3{\times}3$ convolutions.
This outperforms FastInst's approach of directly decoding at the lower
resolution by $+2.2$\,AP at $\nicefrac{1}{16}$ and $+6.6$\,AP at
$\nicefrac{1}{32}$ on COCO \texttt{val2017}
(\cref{fig:token_ablation}), demonstrating that retaining
full-resolution features before pooling and then restoring detail via
the refinement head is far more effective than simply operating at a
coarser scale.
Within this design space we adopt PF\,$\nicefrac{1}{16}$ as the
headline operating point: it matches FastInst's accuracy on both
datasets and underlies the speedups in
\cref{sec:exp:decspeedup,sec:exp:e2e}.
Pushing further to PF\,$\nicefrac{1}{32}$ shows clear diminishing
returns: it removes far fewer tokens in absolute terms ($\sim$6k vs.\
the $\sim$25k saved by the $\nicefrac{1}{8}{\to}\nicefrac{1}{16}$ step
at CS) at a cost of $1.4$\,AP on COCO and $0.8$\,AP on Cityscapes
\texttt{test}.

\paragraph{Panoptic segmentation.}
On Cityscapes panoptic segmentation (\cref{tab:panoptic}), CompactInst
achieves $52.0$\,PQ ($-0.6$ vs.\ FastInst's $52.6$,
${\sim}1\%$ relative), with a similar pattern across things and stuff
categories.
Instance segmentation quality within the panoptic framework
(AP$_\text{Th}$) is slightly higher for CompactInst ($23.1$
vs.\ $22.7$), while mIoU drops by $0.9$.
Given the $2.6{\times}$ model-level speedup on Orin~Nano
(\cref{fig:e2e-3gpu}), CompactInst is the rather preferred operating point
for panoptic segmentation on bandwidth-limited edge hardware.

\begin{table}[!htbp]
\centering
\caption{Panoptic segmentation on Cityscapes \texttt{val}.
  CompactInst incurs ${\sim}1\%$ relative PQ loss while delivering
  substantial latency savings (\cref{sec:exp:e2e}).}
\label{tab:panoptic}
\footnotesize
\renewcommand{\arraystretch}{1.15}
\begin{tabular}{@{}l ccccc@{}}
\hline
Method & PQ & PQ$_\text{Th}$ & PQ$_\text{St}$ & AP$_\text{Th}$ & mIoU \\
\hline
FastInst~\cite{FastInst}
  & \textbf{52.6} & \textbf{41.7} & \textbf{60.6} & 22.7 & \textbf{69.6} \\
CompactInst (ours)
  & 52.0 & 40.9 & 60.0 & \textbf{23.1} & 68.7 \\
\hline
\end{tabular}
\end{table}

\paragraph{Generalization to Mask2Former and comparison with PEM.}
To test whether the recipe transfers beyond FastInst's dual-path
decoder and compare it with the closest decoder-side token-reduction
prior, we evaluate three matched single-path Mask2Former variants
($\nicefrac{1}{8}$ pixel features, three decoder layers): the baseline
decoder, our pooling to $\nicefrac{1}{16}$ plus refinement head, and
PEM's prototype-based cross-attention~\cite{PEM}
(\cref{sec:related}).
\Cref{tab:m2f} reports Cityscapes accuracy (R-50) and TensorRT FP16
latency (R-18 backbone, no post-processing) on Orin~Nano at
$1024{\times}2048$, measured as medians of five runs under locked GPU
and memory clocks (full protocol in the supplementary).
Our recipe leaves test AP unchanged ($21.9$ vs.\ $21.9$) and cuts
decoder latency by $61\%$ (model $-35\%$) despite adding
$231\%$ decoder FLOPs: the refinement head, a fixed compute-dense
addition, alone accounts for $58.1$ of the $64.0$\,GFLOPs ($91\%$).
It dominates the small M2F decoder's FLOP count while the pooled
tokens dominate its bytes.
Under the matched hardware-counter DRAM profile at Full HD, total decoder traffic
is $4.90$\,GB for the baseline, $3.62$\,GB for PEM, and $1.81$\,GB for
our design.
PEM thus reduces total decoder DRAM traffic by $26\%$, substantially
less than our $63\%$ reduction, but its $32\%$ FLOP reduction does not
translate into lower deployment latency at either the decoder or model
level, at a
cost of $2.8$/$2.4$ val/test AP ($1.7$/$0.3$ with its native batch-32
recipe).
In this matched deployment comparison, FLOPs alone do not predict edge
latency.

\begin{table}[h]
  \centering
  \caption{%
    Generalization: matched single-path Mask2Former
    ($\nicefrac{1}{8}$ scale, three layers) on Cityscapes; columns
    differ only in the decoder (AP: R-50, val\,/\,test).
    Latency/FLOPs: TensorRT FP16, decoder / full model (no
    post-processing, R-18), Orin~Nano, $1024{\times}2048$.
    $\ddagger$: PEM's native batch-32 recipe (same architecture/steps,
    $2{\times}$ training samples/compute).%
  }
  \label{tab:m2f}
  \footnotesize
  \renewcommand{\arraystretch}{1.15}
  \setlength{\tabcolsep}{4pt}
  \begin{tabular}{lcccc}
    \hline
    M2F decoder & Baseline & +pool+refine (ours) & PEM~\cite{PEM} & PEM$^\ddagger$\\
    \hline
    AP val\,/\,test        & 25.7\,/\,21.9 & 25.5\,/\,21.9 & 22.9\,/\,19.5 & 24.0\,/\,21.6\\
    AP$_{50}$ val\,/\,test & 51.3\,/\,46.1 & 50.8\,/\,45.7 & 44.5\,/\,41.3 & 48.0\,/\,45.0\\
    \hline
    Decoder FLOPs [G]    & 19.3  & 64.0 (\textcolor{badred}{$+231\%$}) & \multicolumn{2}{c}{13.2 (\textcolor{goodgreen}{$-32\%$})}\\
    Decoder latency [ms] & 98.5 & 38.1 (\textcolor{goodgreen}{$-61\%$})  & \multicolumn{2}{c}{111.0 (\textcolor{badred}{$+13\%$})}\\
    Model FLOPs [G]      & 143.0 & 187.6 (\textcolor{badred}{$+31\%$}) & \multicolumn{2}{c}{136.8 (\textcolor{goodgreen}{$-4\%$})}\\
    Model latency [ms]   & 126.6 & 82.1 (\textcolor{goodgreen}{$-35\%$}) & \multicolumn{2}{c}{156.8 (\textcolor{badred}{$+24\%$})}\\
    \hline
  \end{tabular}
\end{table}

\section{Conclusion}
\label{sec:conclusion}

We have shown that transformer decoders in query-based segmentation
models are limited by memory bandwidth under deployment conditions
(TensorRT FP16 with tensor cores). The decoder contains compute- and
bandwidth-bound kernels under both precisions, but the balance shifts
strongly under FP16 tensor-core execution: the share of time above the
appropriate per-kernel ridge falls from $55.7\%$ to $9.3\%$, while
non-GEMM kernels grow from $36.9\%$ to $53.3\%$ of decoder time.
Measured using hardware DRAM counters, the CUDA-only aggregates of the
standard FastInst and Mask2Former decoders reach only $2.5\%$ to $6\%$
of the CUDA ridge yet account for $53\%$ to $70\%$ of decoder time and
$61\%$ to $71\%$ of decoder bytes. The traffic comes from weight-free
kernels that stream pixel-token activations between GEMMs. The largest
individual kernels sustain up to $90\%$ of nominal peak DRAM bandwidth,
effectively saturating the memory bus.
The M$\nicefrac{1}{4}$ variant provides further deployment-level evidence:
$2.57{\times}$ more decoder FLOPs than FastInst yet $1.74{\times}$
faster on Orin~Nano. This reordering is consistent with the measured
traffic counters, which show that bytes moved, rather than operations
performed, govern execution time in this bandwidth-limited regime.
CompactInst translates the diagnosis into deployment latency:
removing $64\%$ of decoder DRAM bytes yields up to $2.62{\times}$
model-level speedup on Orin~Nano at matched Cityscapes AP and
${\sim}1\%$ relative PQ loss on panoptic segmentation. In our matched
single-path Mask2Former ablation, the same recipe cuts decoder latency
by $61\%$ at unchanged test AP, whereas PEM's prototype-based
cross-attention~\cite{PEM} reduces decoder FLOPs by $32\%$ but increases
latency relative to the baseline.

\paragraph{Limitations and future work.}
Uniform pooling is the simplest byte-reducing lever. A natural next
step is using more sophisticated token reduction that targets bytes moved
rather than decoder FLOPs and is ideally content-aware, so that the
small-object cost observed on COCO also disappears.
Second, post-processing is itself bandwidth-bound and remains another
bottleneck for practical deployment at high resolution. Making it
deployment-efficient remains open and is an interesting direction for future research.

\bibliography{egbib}

\begin{thebibliography}{39}
\providecommand{\natexlab}[1]{#1}
\providecommand{\url}[1]{\texttt{#1}}
\expandafter\ifx\csname urlstyle\endcsname\relax
  \providecommand{\doi}[1]{doi: #1}\else
  \providecommand{\doi}{doi: \begingroup \urlstyle{rm}\Url}\fi

\bibitem[Bolya et~al.(2023)Bolya, Fu, Dai, Zhang, Feichtenhofer, and
  Hoffman]{ToMe}
Daniel Bolya, Cheng-Yang Fu, Xiaoliang Dai, Peizhao Zhang, Christoph
  Feichtenhofer, and Judy Hoffman.
\newblock Token merging: Your {ViT} but faster.
\newblock In \emph{ICLR}, 2023.

\bibitem[Caesar et~al.(2020)Caesar, Bankiti, Lang, Vora, Liong, Xu, Krishnan,
  Pan, Baldan, and Beijbom]{nuscenes2019}
Holger Caesar, Varun Bankiti, Alex~H. Lang, Sourabh Vora, Venice~Erin Liong,
  Qiang Xu, Anush Krishnan, Yu~Pan, Giancarlo Baldan, and Oscar Beijbom.
\newblock nu{S}cenes: A multimodal dataset for autonomous driving.
\newblock In \emph{CVPR}, 2020.

\bibitem[Carion et~al.(2020)Carion, Massa, Synnaeve, Usunier, Kirillov, and
  Zagoruyko]{DETR}
Nicolas Carion, Francisco Massa, Gabriel Synnaeve, Nicolas Usunier, Alexander
  Kirillov, and Sergey Zagoruyko.
\newblock End-to-end object detection with transformers.
\newblock In \emph{ECCV}, 2020.

\bibitem[Cavagnero et~al.(2024)Cavagnero, Rosi, Cuttano, Pistilli, Ciccone,
  Averta, and Cermelli]{PEM}
Niccol{\`o} Cavagnero, Gabriele Rosi, Claudia Cuttano, Francesca Pistilli,
  Marco Ciccone, Giuseppe Averta, and Fabio Cermelli.
\newblock {PEM}: Prototype-based efficient {MaskFormer} for image segmentation.
\newblock In \emph{CVPR}, 2024.

\bibitem[Cheng et~al.(2021)Cheng, Schwing, and Kirillov]{MaskFormer}
Bowen Cheng, Alexander~G. Schwing, and Alexander Kirillov.
\newblock Per-pixel classification is not all you need for semantic
  segmentation.
\newblock In \emph{NeurIPS}, 2021.

\bibitem[Cheng et~al.(2022{\natexlab{a}})Cheng, Misra, Schwing, Kirillov, and
  Girdhar]{Mask2Former}
Bowen Cheng, Ishan Misra, Alexander~G. Schwing, Alexander Kirillov, and Rohit
  Girdhar.
\newblock Masked-attention mask transformer for universal image segmentation.
\newblock In \emph{CVPR}, 2022{\natexlab{a}}.

\bibitem[Cheng et~al.(2022{\natexlab{b}})Cheng, Wang, Chen, Zhang, Zhang,
  Huang, Zhang, and Liu]{SparseInst}
Tianheng Cheng, Xinggang Wang, Shaoyu Chen, Wenqiang Zhang, Qian Zhang, Chang
  Huang, Zhaoxiang Zhang, and Wenyu Liu.
\newblock Sparse instance activation for real-time instance segmentation.
\newblock In \emph{CVPR}, 2022{\natexlab{b}}.

\bibitem[Cordts et~al.(2016)Cordts, Omran, Ramos, Rehfeld, Enzweiler, Benenson,
  Franke, Roth, and Schiele]{Cordts2016Cityscapes}
Marius Cordts, Mohamed Omran, Sebastian Ramos, Timo Rehfeld, Markus Enzweiler,
  Rodrigo Benenson, Uwe Franke, Stefan Roth, and Bernt Schiele.
\newblock The {Cityscapes} dataset for semantic urban scene understanding.
\newblock In \emph{CVPR}, 2016.

\bibitem[Dao et~al.(2022)Dao, Fu, Ermon, Rudra, and
  R{\'e}]{Dao2022FlashAttention}
Tri Dao, Daniel~Y. Fu, Stefano Ermon, Atri Rudra, and Christopher R{\'e}.
\newblock {FlashAttention}: Fast and memory-efficient exact attention with
  {IO}-awareness.
\newblock In \emph{NeurIPS}, 2022.

\bibitem[Deng et~al.(2009)Deng, Dong, Socher, Li, Li, and Fei-Fei]{ImageNet}
Jia Deng, Wei Dong, Richard Socher, Li-Jia Li, Kai Li, and Li~Fei-Fei.
\newblock Image{N}et: A large-scale hierarchical image database.
\newblock In \emph{CVPR}, 2009.

\bibitem[Gholami et~al.(2024)Gholami, Yao, Kim, Hooper, Mahoney, and
  Keutzer]{MemoryWall}
Amir Gholami, Zhewei Yao, Sehoon Kim, Coleman Hooper, Michael~W. Mahoney, and
  Kurt Keutzer.
\newblock {AI} and memory wall.
\newblock \emph{arXiv preprint arXiv:2403.14123}, 2024.

\bibitem[Goyal et~al.(2017)Goyal, Doll{\'a}r, Girshick, Noordhuis, Wesolowski,
  Kyrola, Tulloch, Jia, and He]{Goyal2017LinearScaling}
Priya Goyal, Piotr Doll{\'a}r, Ross Girshick, Pieter Noordhuis, Lukasz
  Wesolowski, Aapo Kyrola, Andrew Tulloch, Yangqing Jia, and Kaiming He.
\newblock Accurate, large minibatch {SGD}: Training {ImageNet} in 1 hour.
\newblock \emph{arXiv preprint arXiv:1706.02677}, 2017.

\bibitem[He et~al.(2023)He, Li, Geng, and Xie]{FastInst}
Junjie He, Pengyu Li, Yifeng Geng, and Xuansong Xie.
\newblock {FastInst}: A simple query-based model for real-time instance
  segmentation.
\newblock In \emph{CVPR}, 2023.

\bibitem[He et~al.(2016)He, Zhang, Ren, and Sun]{ResNet}
Kaiming He, Xiangyu Zhang, Shaoqing Ren, and Jian Sun.
\newblock Deep residual learning for image recognition.
\newblock In \emph{CVPR}, 2016.

\bibitem[He et~al.(2017)He, Gkioxari, Doll{\'a}r, and Girshick]{MaskRCNN}
Kaiming He, Georgia Gkioxari, Piotr Doll{\'a}r, and Ross Girshick.
\newblock Mask {R-CNN}.
\newblock In \emph{ICCV}, 2017.

\bibitem[Ivanov et~al.(2021)Ivanov, Dryden, Ben-Nun, Li, and
  Hoefler]{Ivanov2021DataMovement}
Andrei Ivanov, Nikoli Dryden, Tal Ben-Nun, Shigang Li, and Torsten Hoefler.
\newblock Data movement is all you need: A case study on optimizing
  transformers.
\newblock In \emph{MLSys}, 2021.

\bibitem[Kao et~al.(2023)Kao, Subramanian, Agrawal, Yazdanbakhsh, and
  Krishna]{Kao2023FLAT}
Sheng-Chun Kao, Suvinay Subramanian, Gaurav Agrawal, Amir Yazdanbakhsh, and
  Tushar Krishna.
\newblock {FLAT}: An optimized dataflow for mitigating attention bottlenecks.
\newblock In \emph{ASPLOS}, 2023.

\bibitem[Liang et~al.(2022)Liang, Ge, Tong, Song, Wang, and Xie]{EViT}
Youwei Liang, Chongjian Ge, Zhan Tong, Yibing Song, Jue Wang, and Pengtao Xie.
\newblock Not all patches are what you need: Expediting vision transformers via
  token reorganizations.
\newblock In \emph{ICLR}, 2022.

\bibitem[Lin et~al.(2014)Lin, Maire, Belongie, Hays, Perona, Ramanan,
  Doll{\'a}r, and Zitnick]{COCO}
Tsung-Yi Lin, Michael Maire, Serge Belongie, James Hays, Pietro Perona, Deva
  Ramanan, Piotr Doll{\'a}r, and C.~Lawrence Zitnick.
\newblock Microsoft {COCO}: Common objects in context.
\newblock In \emph{ECCV}, 2014.

\bibitem[Liu et~al.(2023)Liu, Peng, Zheng, Yang, Hu, and
  Yuan]{Liu2023EfficientViT}
Xinyu Liu, Houwen Peng, Ningxin Zheng, Yuqing Yang, Han Hu, and Yixuan Yuan.
\newblock {EfficientViT}: Memory efficient vision transformer with cascaded
  group attention.
\newblock In \emph{CVPR}, 2023.

\bibitem[Liu et~al.(2024)Liu, Zhang, Khaki, Yang, Tang, Xu, Keutzer, and
  Han]{SparseRefine}
Zhijian Liu, Zhuoyang Zhang, Samir Khaki, Shang Yang, Haotian Tang, Chenfeng
  Xu, Kurt Keutzer, and Song Han.
\newblock Sparse refinement for efficient high-resolution semantic
  segmentation.
\newblock In \emph{ECCV}, 2024.

\bibitem[Loshchilov and Hutter(2019)]{AdamW}
Ilya Loshchilov and Frank Hutter.
\newblock Decoupled weight decay regularization.
\newblock In \emph{ICLR}, 2019.

\bibitem[Ma et~al.(2018)Ma, Zhang, Zheng, and Sun]{ShuffleNetV2}
Ningning Ma, Xiangyu Zhang, Hai-Tao Zheng, and Jian Sun.
\newblock {ShuffleNet V2}: Practical guidelines for efficient {CNN}
  architecture design.
\newblock In \emph{ECCV}, 2018.

\bibitem[Nayal et~al.(2023)Nayal, Yavuz, Henriques, and G{\"u}ney]{RbA}
Nazir Nayal, M{\i}sra Yavuz, Jo{\~a}o~F. Henriques, and Fatma G{\"u}ney.
\newblock {RbA}: Segmenting unknown regions rejected by all.
\newblock In \emph{ICCV}, 2023.

\bibitem[Oh et~al.(2026)Oh, Chen, Yun, Ni, Huang, Das, Jang, and Imani]{QUILL}
Hyunwoo Oh, Hanning Chen, Sanggeon Yun, Yang Ni, Wenjun Huang, Tamoghno Das,
  Suyeon Jang, and Mohsen Imani.
\newblock {QUILL}: An algorithm--architecture co-design for cache-local
  deformable attention.
\newblock In \emph{DATE}, 2026.

\bibitem[Pan et~al.(2022)Pan, Cai, and Zhuang]{Pan2022LITv2HiLo}
Zizheng Pan, Jianfei Cai, and Bohan Zhuang.
\newblock Fast vision transformers with {HiLo} attention.
\newblock In \emph{NeurIPS}, 2022.

\bibitem[Rai et~al.(2024)Rai, Cermelli, Fontanel, Masone, and
  Caputo]{Mask2Anomaly}
Shyam~Nandan Rai, Fabio Cermelli, Dario Fontanel, Carlo Masone, and Barbara
  Caputo.
\newblock {Mask2Anomaly}: Mask transformer for universal open-set segmentation.
\newblock In \emph{CVPR}, 2024.

\bibitem[Rao et~al.(2021)Rao, Zhao, Liu, Lu, Zhou, and Hsieh]{DynamicViT}
Yongming Rao, Wenliang Zhao, Benlin Liu, Jiwen Lu, Jie Zhou, and Cho-Jui Hsieh.
\newblock {DynamicViT}: Efficient vision transformers with dynamic token
  sparsification.
\newblock In \emph{NeurIPS}, 2021.

\bibitem[Ryoo et~al.(2022)Ryoo, Piergiovanni, Arnab, Dehghani, and
  Angelova]{TokenLearner}
Michael~S. Ryoo, AJ~Piergiovanni, Anurag Arnab, Mostafa Dehghani, and Anelia
  Angelova.
\newblock {TokenLearner}: What can 8 learned tokens do for images and videos?
\newblock \emph{IEEE TPAMI}, 44\penalty0 (12):\penalty0 9942--9952, 2022.
\newblock \doi{10.1109/TPAMI.2021.3140060}.

\bibitem[Segu et~al.(2025)Segu, Gazulla, Xian, Gool, and Tombari]{MOBIUS}
Mattia Segu, Marta~Tintore Gazulla, Yongqin Xian, Luc~Van Gool, and Federico
  Tombari.
\newblock {MOBIUS}: Big-to-mobile universal instance segmentation via
  multi-modal bottleneck fusion and calibrated decoder pruning.
\newblock In \emph{ICCV}, 2025.

\bibitem[Sun et~al.(2020)Sun, Kretzschmar, Dotiwalla, Chouard, Patnaik, Tsui,
  Guo, Zhou, Chai, Caine, Vasudevan, Han, Ngiam, Zhao, Timofeev, Ettinger,
  Krivokon, Gao, Joshi, Zhang, Shlens, Chen, and Anguelov]{Sun_2020_CVPR}
Pei Sun, Henrik Kretzschmar, Xerxes Dotiwalla, Aurelien Chouard, Vijaysai
  Patnaik, Paul Tsui, James Guo, Yin Zhou, Yuning Chai, Benjamin Caine, Vijay
  Vasudevan, Wei Han, Jiquan Ngiam, Hang Zhao, Aleksei Timofeev, Scott
  Ettinger, Maxim Krivokon, Amy Gao, Aditya Joshi, Yu~Zhang, Jonathon Shlens,
  Zhifeng Chen, and Dragomir Anguelov.
\newblock Scalability in perception for autonomous driving: {Waymo Open
  Dataset}.
\newblock In \emph{CVPR}, 2020.

\bibitem[Tian et~al.(2020)Tian, Shen, and Chen]{CondInst}
Zhi Tian, Chunhua Shen, and Hao Chen.
\newblock Conditional convolutions for instance segmentation.
\newblock In \emph{ECCV}, 2020.

\bibitem[Vaswani et~al.(2017)Vaswani, Shazeer, Parmar, Uszkoreit, Jones, Gomez,
  Kaiser, and Polosukhin]{Transformer}
Ashish Vaswani, Noam Shazeer, Niki Parmar, Jakob Uszkoreit, Llion Jones,
  Aidan~N. Gomez, {\L}ukasz Kaiser, and Illia Polosukhin.
\newblock Attention is all you need.
\newblock In \emph{NeurIPS}, 2017.

\bibitem[Wang et~al.(2020)Wang, Zhang, Kong, Li, and Shen]{SOLOv2}
Xinlong Wang, Rufeng Zhang, Tao Kong, Lei Li, and Chunhua Shen.
\newblock {SOLOv2}: Dynamic and fast instance segmentation.
\newblock In \emph{NeurIPS}, 2020.

\bibitem[Williams et~al.(2009)Williams, Waterman, and
  Patterson]{Williams2009Roofline}
Samuel Williams, Andrew Waterman, and David Patterson.
\newblock Roofline: An insightful visual performance model for multicore
  architectures.
\newblock \emph{Commun. ACM}, 52\penalty0 (4):\penalty0 65--76, 2009.
\newblock \doi{10.1145/1498765.1498785}.

\bibitem[Wu et~al.(2019)Wu, Kirillov, Massa, Lo, and Girshick]{Detectron2}
Yuxin Wu, Alexander Kirillov, Francisco Massa, Wan-Yen Lo, and Ross Girshick.
\newblock {Detectron2}.
\newblock \url{https://github.com/facebookresearch/detectron2}, 2019.

\bibitem[Yang et~al.(2020)Yang, Kurth, and
  Williams]{Yang2020HierarchicalRoofline}
Charlene Yang, Thorsten Kurth, and Samuel Williams.
\newblock Hierarchical roofline analysis for {GPUs}: Accelerating performance
  optimization for the {NERSC}-9 {Perlmutter} system.
\newblock \emph{Concurrency and Computation: Practice and Experience},
  32\penalty0 (20):\penalty0 e5547, 2020.
\newblock \doi{10.1002/cpe.5547}.

\bibitem[Yu et~al.(2020)Yu, Chen, Wang, Xian, Chen, Liu, Madhavan, and
  Darrell]{Yu_2020_CVPR}
Fisher Yu, Haofeng Chen, Xin Wang, Wenqi Xian, Yingying Chen, Fangchen Liu,
  Vashisht Madhavan, and Trevor Darrell.
\newblock {BDD100K}: A diverse driving dataset for heterogeneous multitask
  learning.
\newblock In \emph{CVPR}, 2020.

\bibitem[Zhou and Shi(2025)]{HyPiDecoder}
Fengzhe Zhou and Humphrey Shi.
\newblock {HyPiDecoder}: Hybrid pixel decoder for efficient segmentation and
  detection.
\newblock In \emph{ICCV}, 2025.

\end{thebibliography}

\end{document}


\maketitle
\vspace{-3em}

\appendix
\renewcommand{\thefigure}{\thesection.\arabic{figure}}
\renewcommand{\thetable}{\thesection.\arabic{table}}

\section{Additional Rooflines}
\label{sec:supp-roofline-a4000}

\begin{figure}[h]
  \centering
  \begin{subfigure}[b]{0.49\linewidth}
    \includegraphics[width=\linewidth]{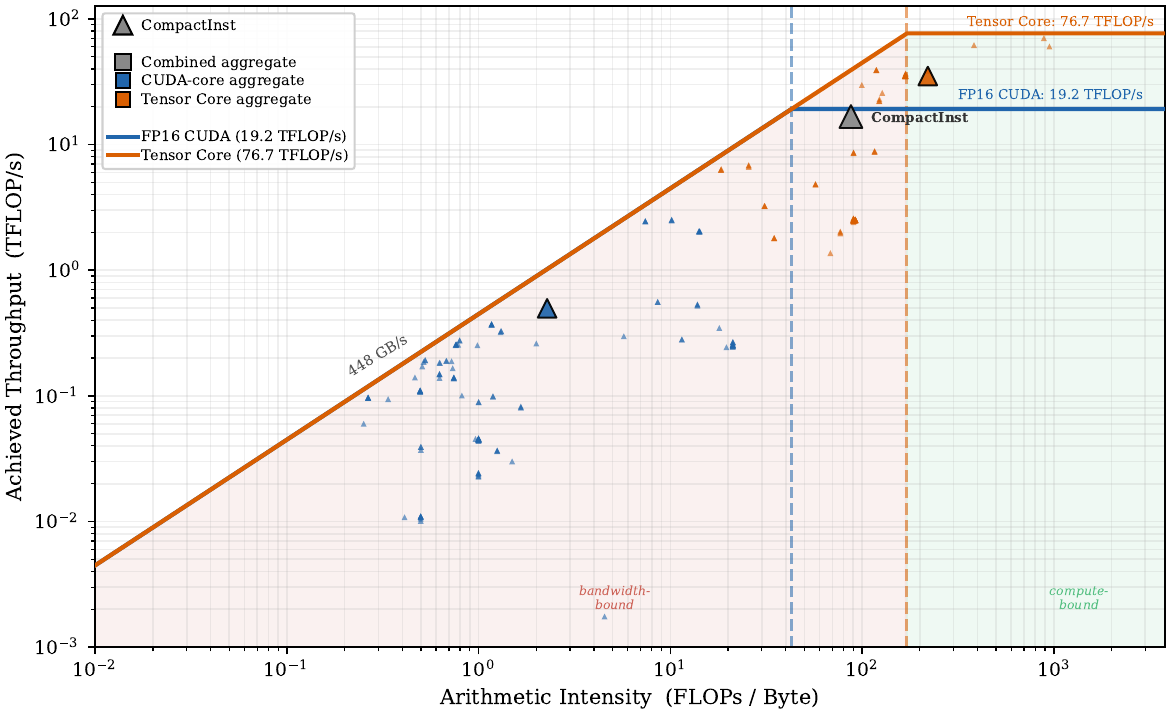}
    \caption{FP16 (tensor cores active)}
  \end{subfigure}\hfill
  \begin{subfigure}[b]{0.49\linewidth}
    \includegraphics[width=\linewidth]{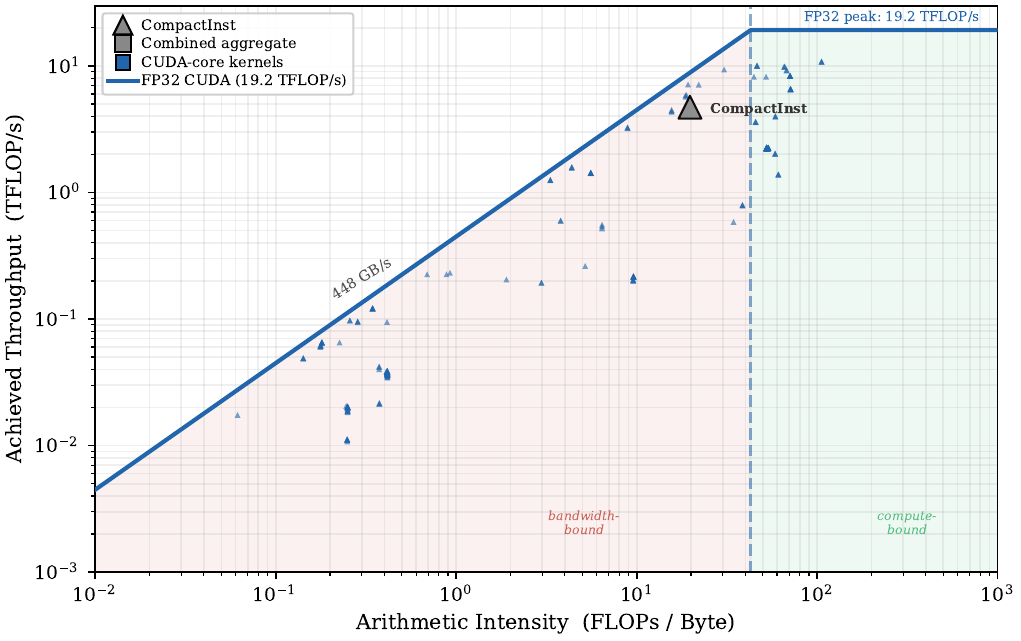}
    \caption{FP32 (tensor cores idle)}
  \end{subfigure}
  \caption{%
    \textbf{CompactInst decoder roofline} on the RTX~A4000 (hardware DRAM
    counters, Full HD $1088{\times}1920$). Orange denotes kernels with
    tensor-core instructions, blue denotes CUDA-only kernels, and large
    markers denote aggregates.%
  }
  \label{fig:supp-a4000-ci}
\end{figure}

\begin{figure}[h]
  \centering
  \begin{subfigure}[b]{0.49\linewidth}
    \includegraphics[width=\linewidth]{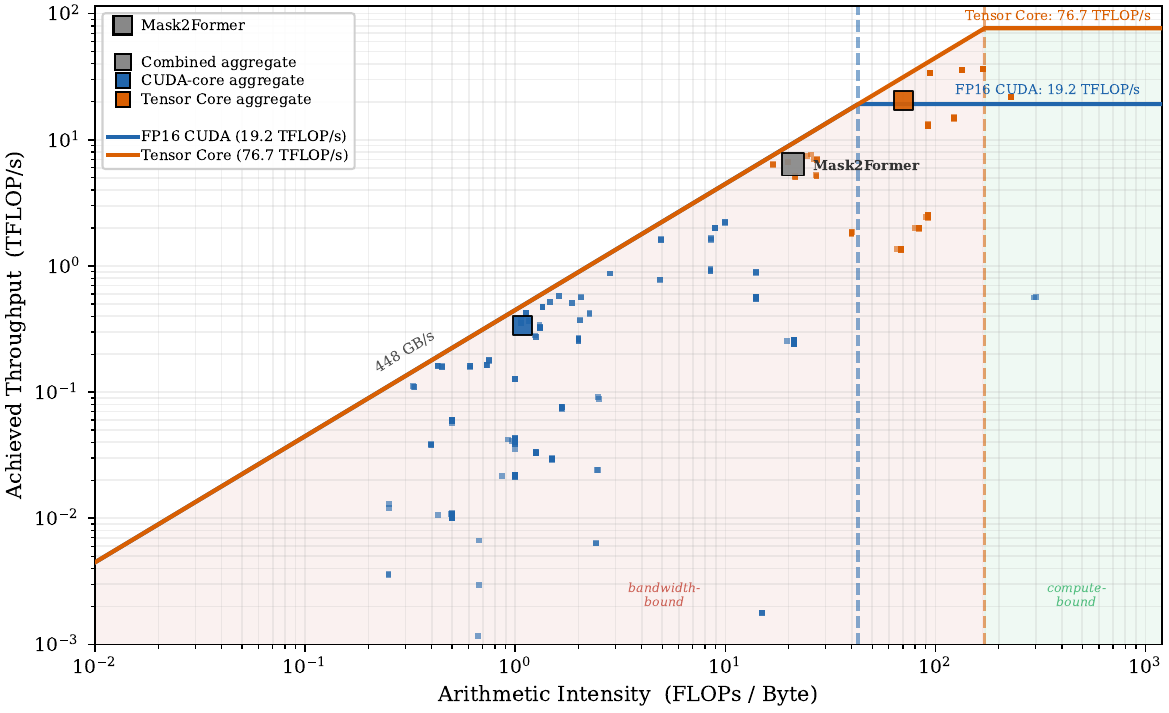}
    \caption{FP16 (tensor cores active)}
  \end{subfigure}\hfill
  \begin{subfigure}[b]{0.49\linewidth}
    \includegraphics[width=\linewidth]{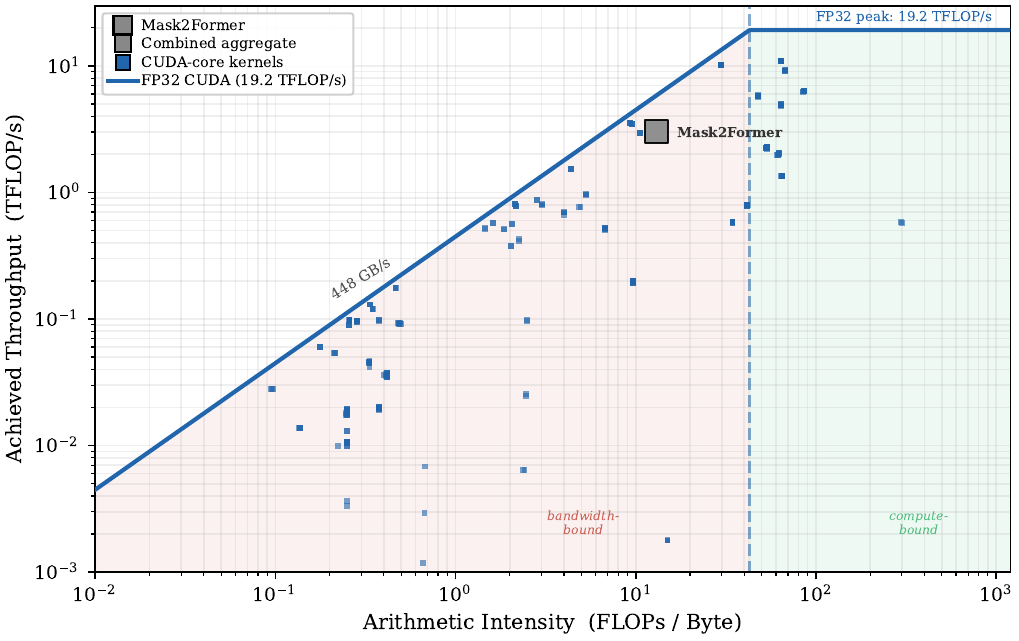}
    \caption{FP32 (tensor cores idle)}
  \end{subfigure}
  \caption{%
    \textbf{Mask2Former decoder roofline} on the RTX~A4000 (hardware DRAM
    counters, Full HD $1088{\times}1920$; same protocol as
    \cref{fig:supp-a4000-ci}). Large markers denote aggregates.%
  }
  \label{fig:supp-a4000-m2f}
\end{figure}

We additionally report FP16 and FP32 rooflines for the CompactInst and
standard Mask2Former decoders (\cref{fig:supp-a4000-ci,fig:supp-a4000-m2f}).
Both use the main paper's RTX~A4000 hardware-counter protocol at Full HD
$1088{\times}1920$ (\texttt{dram\_\_bytes.sum}).
The corresponding aggregate values are reported in the main paper; the
figures below show their per-kernel distributions.

\paragraph{CompactInst.}
Under FP16, CompactInst's CUDA-only activation kernels remain
bandwidth-bound, while the refinement-head convolutions form a
compute-dense tensor-core cluster; FP32 shows the corresponding mixed
regime (\cref{fig:supp-a4000-ci}).

\paragraph{Mask2Former.}
Mask2Former shows the same separation: low-intensity CUDA-only activation
kernels under FP16 and a compute-bound GEMM cluster under FP32
(\cref{fig:supp-a4000-m2f}).

\section{M\texorpdfstring{$\nicefrac{1}{4}$}{1/4} Deployment Latency}
\label{sec:supp-m14-latency}

The M\,$\nicefrac{1}{4}$\,PF\,$\nicefrac{1}{16}$ variant retains
CompactInst's pooled $\nicefrac{1}{16}$ transformer path but refines mask
features at $\nicefrac{1}{4}$ resolution. At Cityscapes resolution this
raises decoder FLOPs to $264.9$\,G, $2.57{\times}$ FastInst's $103.2$\,G;
total model FLOPs rise to $446.5$\,G, $2.17{\times}$ FastInst's
$205.8$\,G. The variant therefore provides a direct counterfactual for
whether arithmetic cost or activation traffic governs latency.

\begin{figure}[H]
  \centering
  \includegraphics[width=\linewidth]{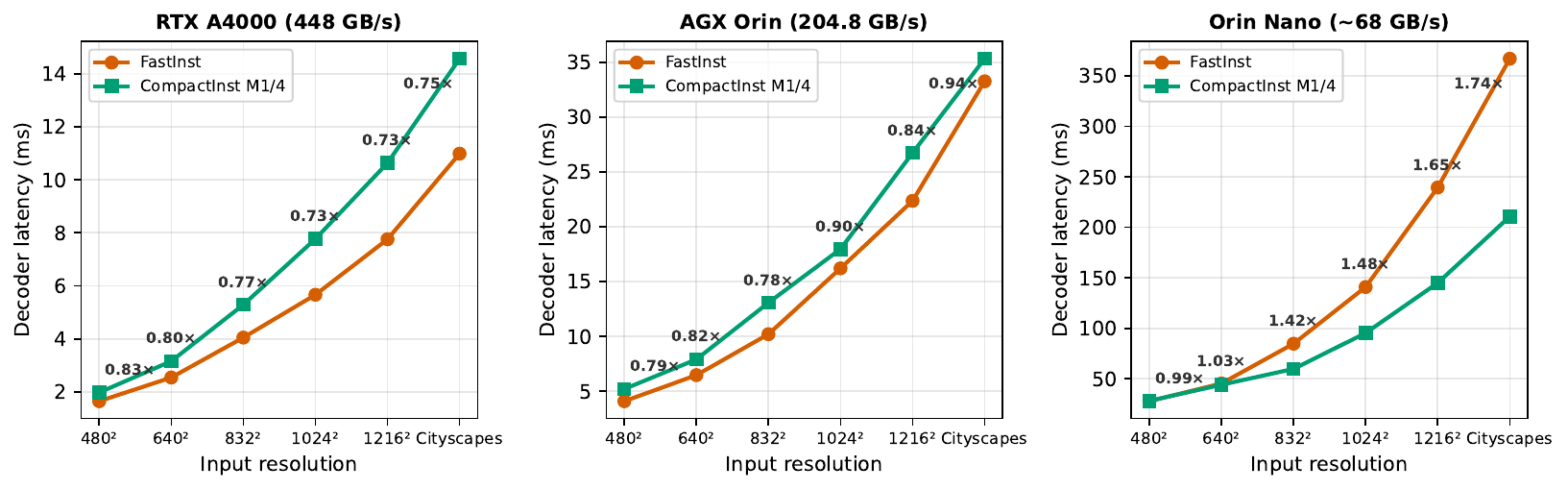}
  \caption{%
    \textbf{Decoder-only TensorRT FP16 latency} across three GPUs.
    Annotations show the FastInst/M\,$\nicefrac{1}{4}$ latency ratio;
    values above $1{\times}$ indicate that M\,$\nicefrac{1}{4}$ is faster.%
  }
  \label{fig:supp-m14-decoder}
\end{figure}

\begin{figure}[H]
  \centering
  \includegraphics[width=\linewidth]{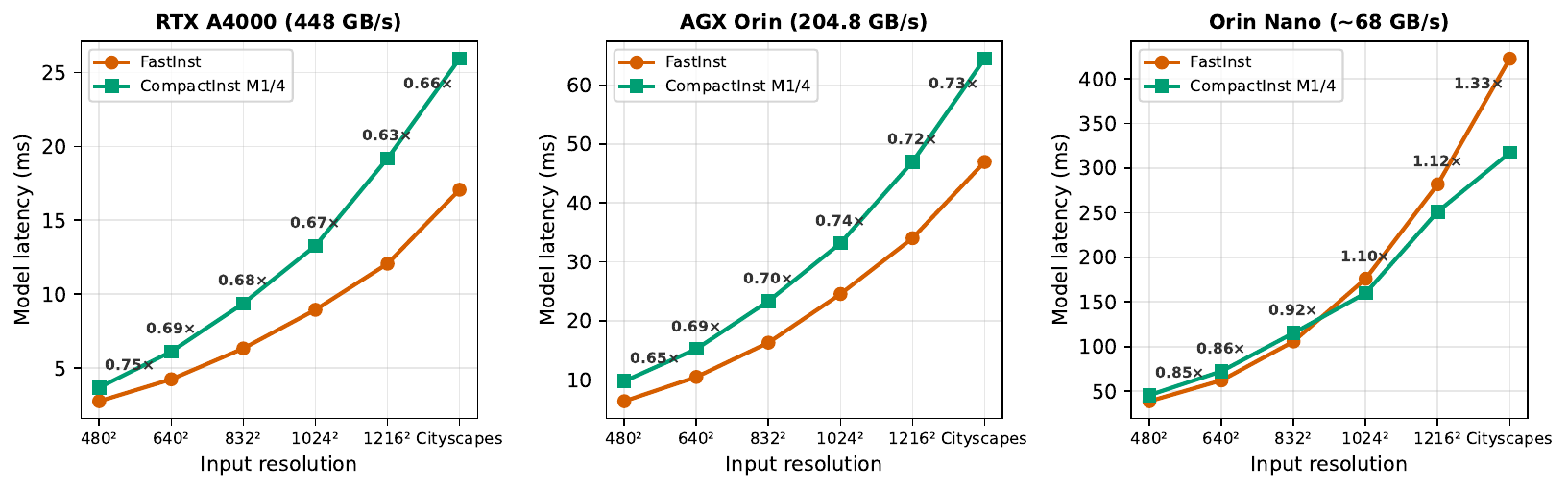}
  \caption{%
    \textbf{Model-level TensorRT FP16 latency} across three GPUs
    (ResNet-18, no post-processing). Annotations show the
    FastInst/M\,$\nicefrac{1}{4}$ latency ratio; values above $1{\times}$
    indicate that M\,$\nicefrac{1}{4}$ is faster.%
  }
  \label{fig:supp-m14-model}
\end{figure}

\Cref{fig:supp-m14-decoder,fig:supp-m14-model} report TensorRT FP16
latency across the three GPUs and six resolutions used in the main paper.
Each annotation is the ratio FastInst/M\,$\nicefrac{1}{4}$: values below
$1{\times}$ denote a slowdown, whereas values above $1{\times}$ denote a
speedup. The variant remains slower on the A4000 and AGX~Orin. On
Orin~Nano, the measured decoder ratio exceeds $1{\times}$ from $640^2$
onward and reaches $1.74{\times}$ at Cityscapes resolution; the measured
model ratio first exceeds $1{\times}$ at $1024^2$ and reaches
$1.33{\times}$. The later model crossover occurs because
M\,$\nicefrac{1}{4}$ also upsamples the FPN one more stage, from
$\nicefrac{1}{8}$ to $\nicefrac{1}{4}$, adding compute outside the pooled
transformer path.

\section{Qualitative Comparison on Cityscapes}
\label{sec:supp-qualitative}

\begin{figure}[p]
  \centering
  \begin{subfigure}[b]{\linewidth}
    \includegraphics[width=\linewidth]{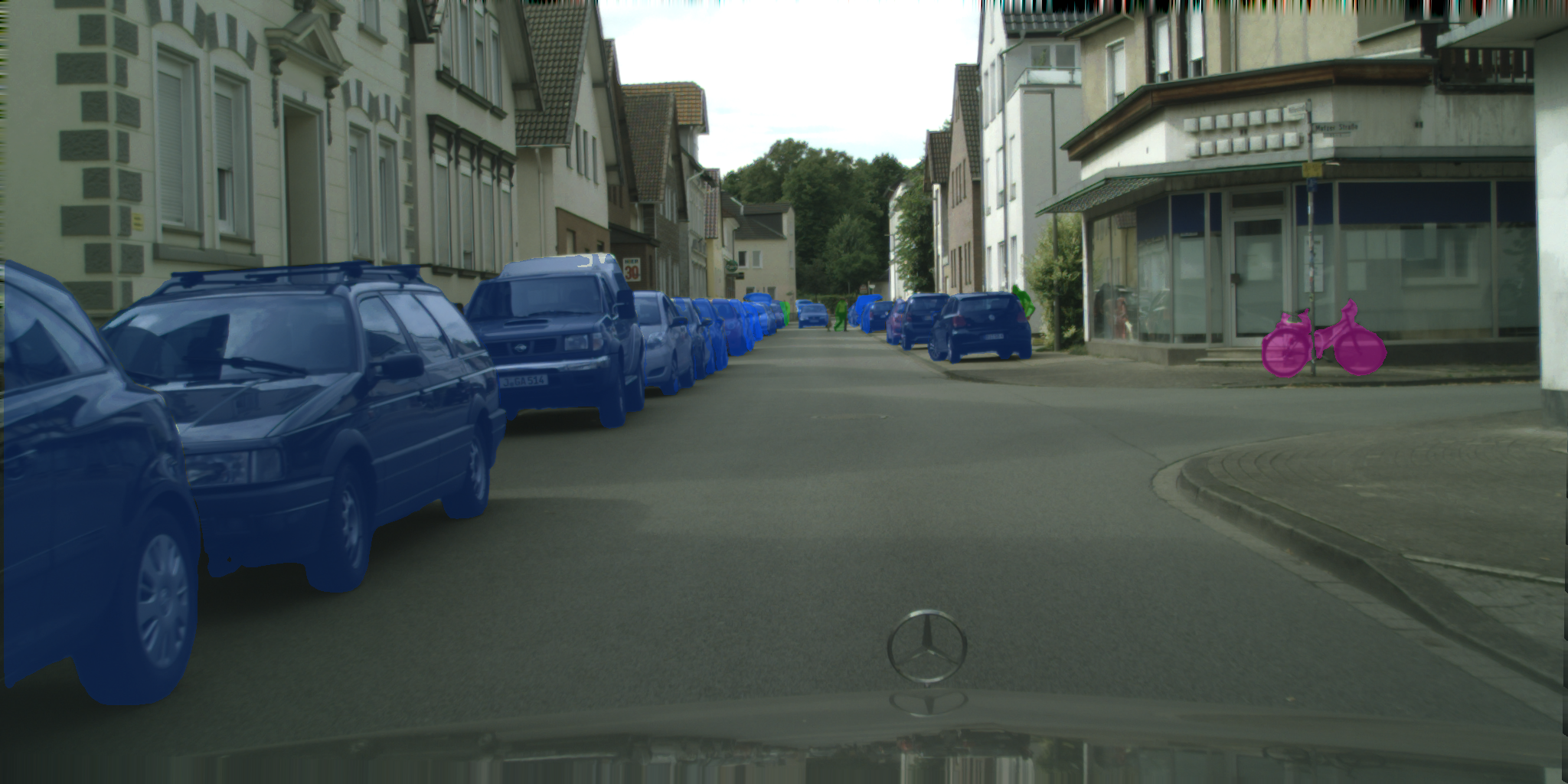}
    \caption{FastInst}
  \end{subfigure}\\[4pt]
  \begin{subfigure}[b]{\linewidth}
    \includegraphics[width=\linewidth]{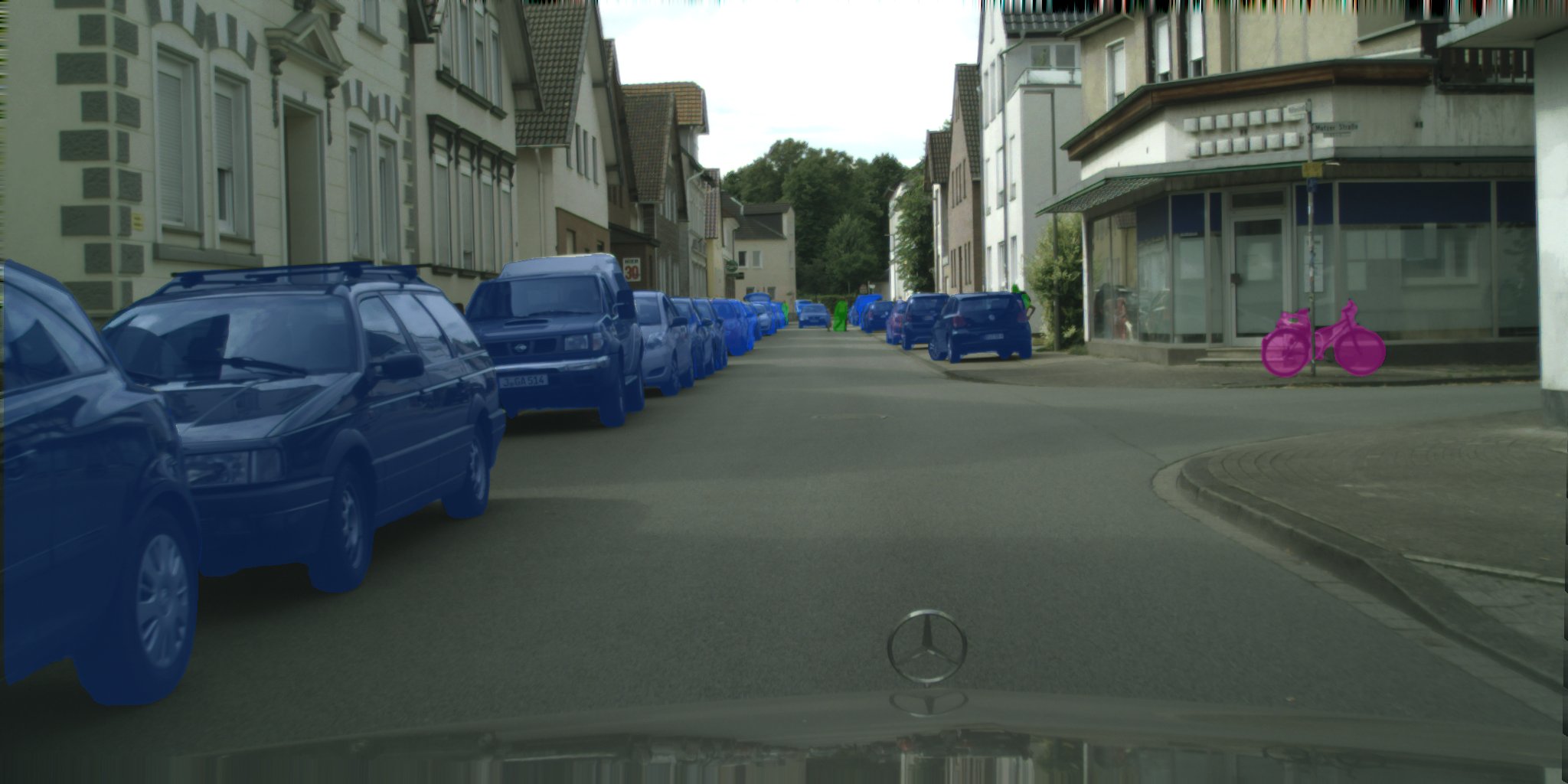}
    \caption{CompactInst (M\,$\nicefrac{1}{8}$\,PF\,$\nicefrac{1}{16}$)}
  \end{subfigure}
  \caption{%
    \textbf{Example~1} (1/2) - FastInst and CompactInst on a
    Cityscapes \texttt{test} image.%
  }
  \label{fig:qual-exp1a}
\end{figure}

\begin{figure}[p]
  \centering
  \begin{subfigure}[b]{\linewidth}
    \includegraphics[width=\linewidth]{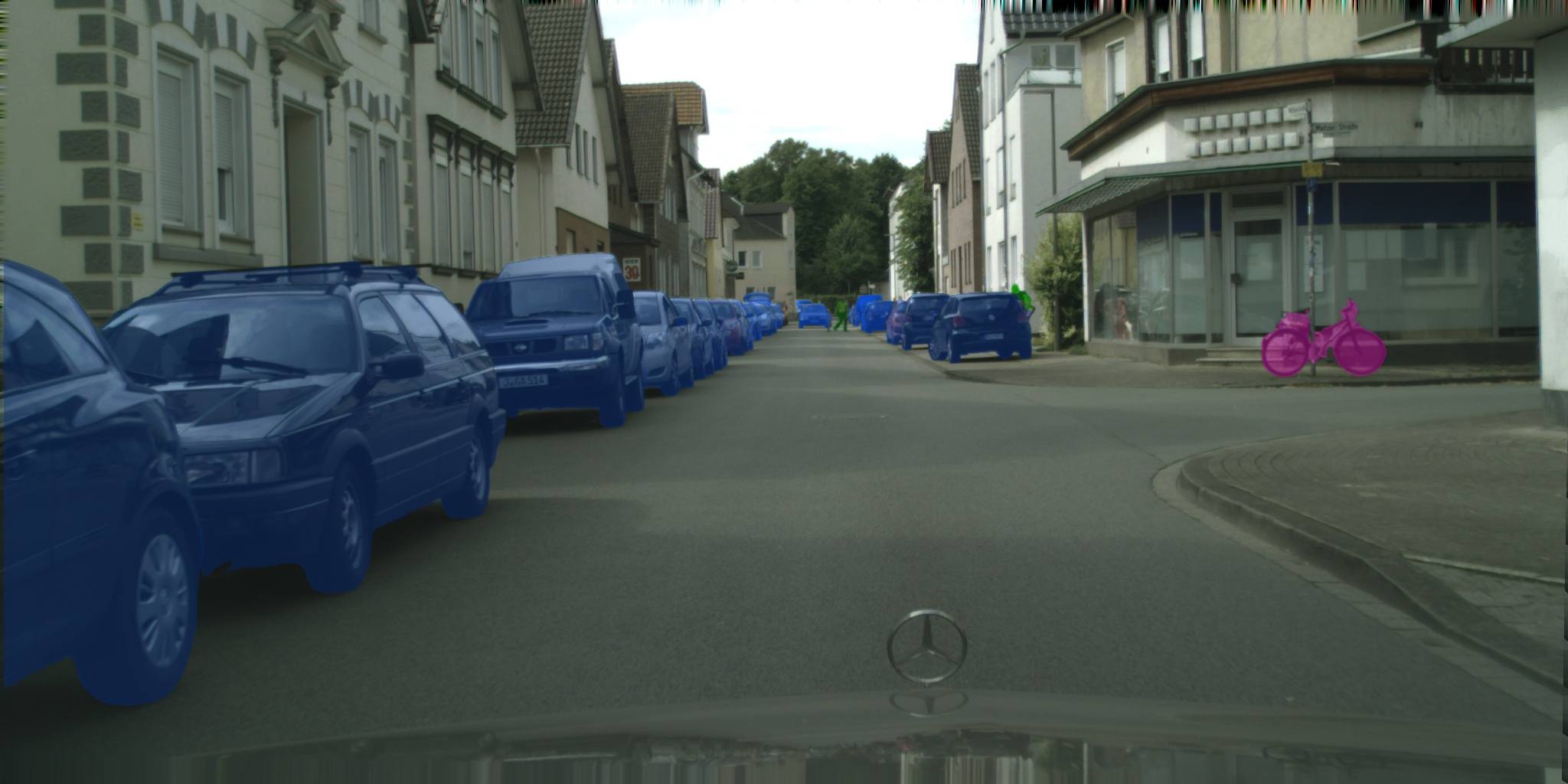}
    \caption{CompactInst w/ M\,$\nicefrac{1}{4}$\,PF\,$\nicefrac{1}{16}$}
  \end{subfigure}\\[4pt]
  \begin{subfigure}[b]{\linewidth}
    \includegraphics[width=\linewidth]{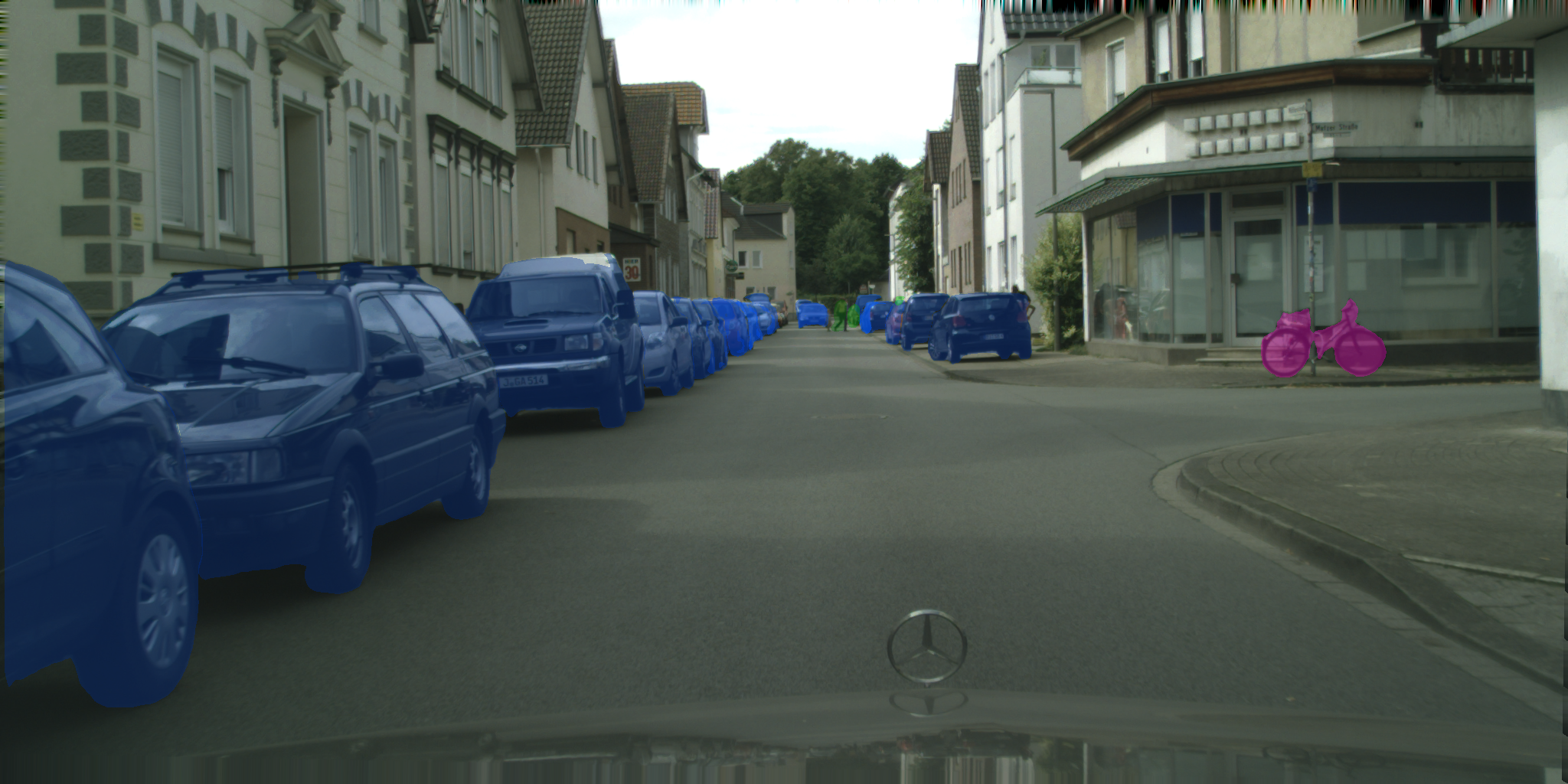}
    \caption{CompactInst w/ PF\,$\nicefrac{1}{32}$}
  \end{subfigure}
  \caption{%
    \textbf{Example~1} (2/2) - CompactInst variants on the same
    Cityscapes \texttt{test} image as \cref{fig:qual-exp1a}.%
  }
  \label{fig:qual-exp1b}
\end{figure}

\begin{figure}[p]
  \centering
  \begin{subfigure}[b]{\linewidth}
    \includegraphics[width=\linewidth]{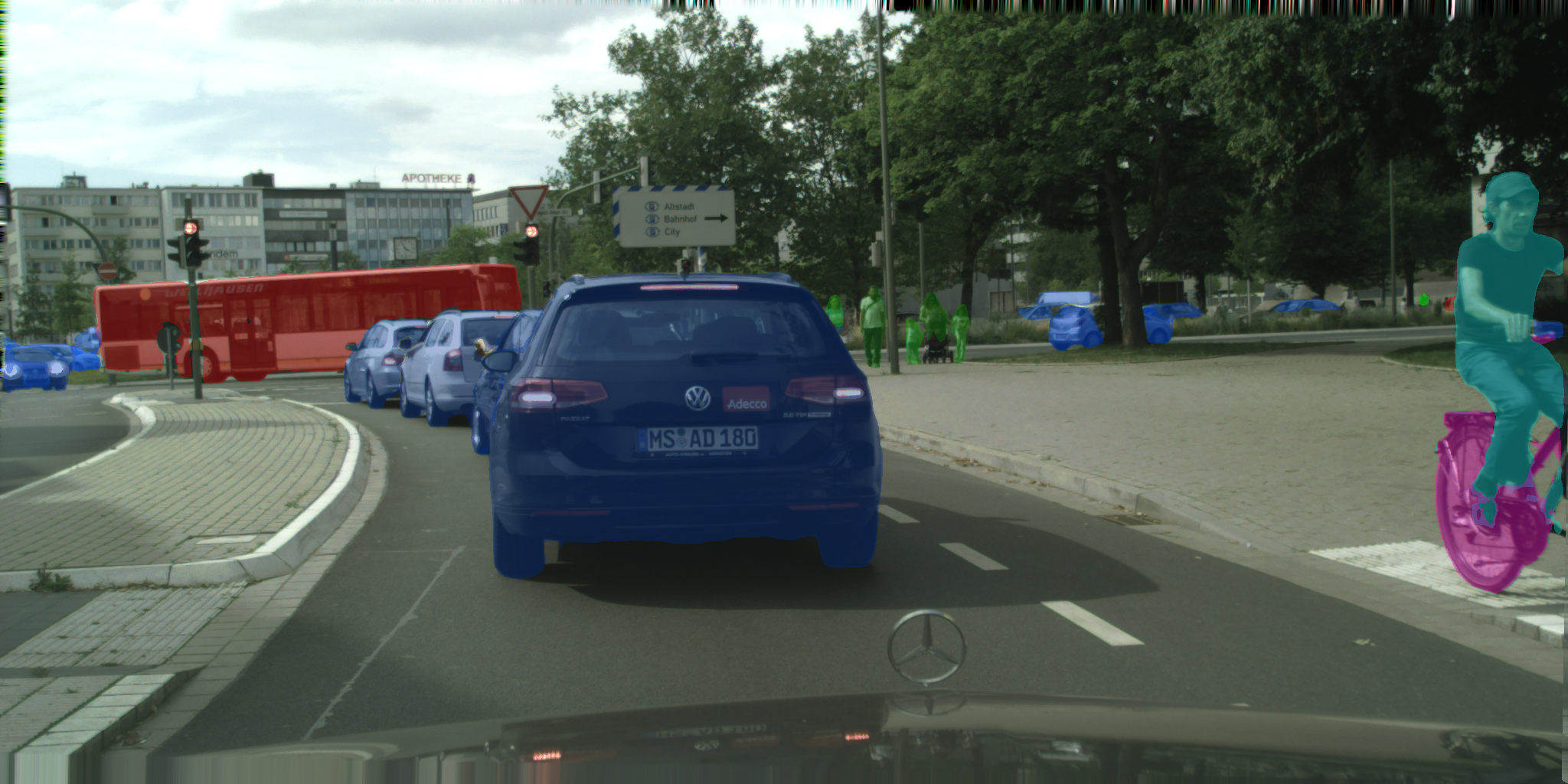}
    \caption{FastInst}
  \end{subfigure}\\[4pt]
  \begin{subfigure}[b]{\linewidth}
    \includegraphics[width=\linewidth]{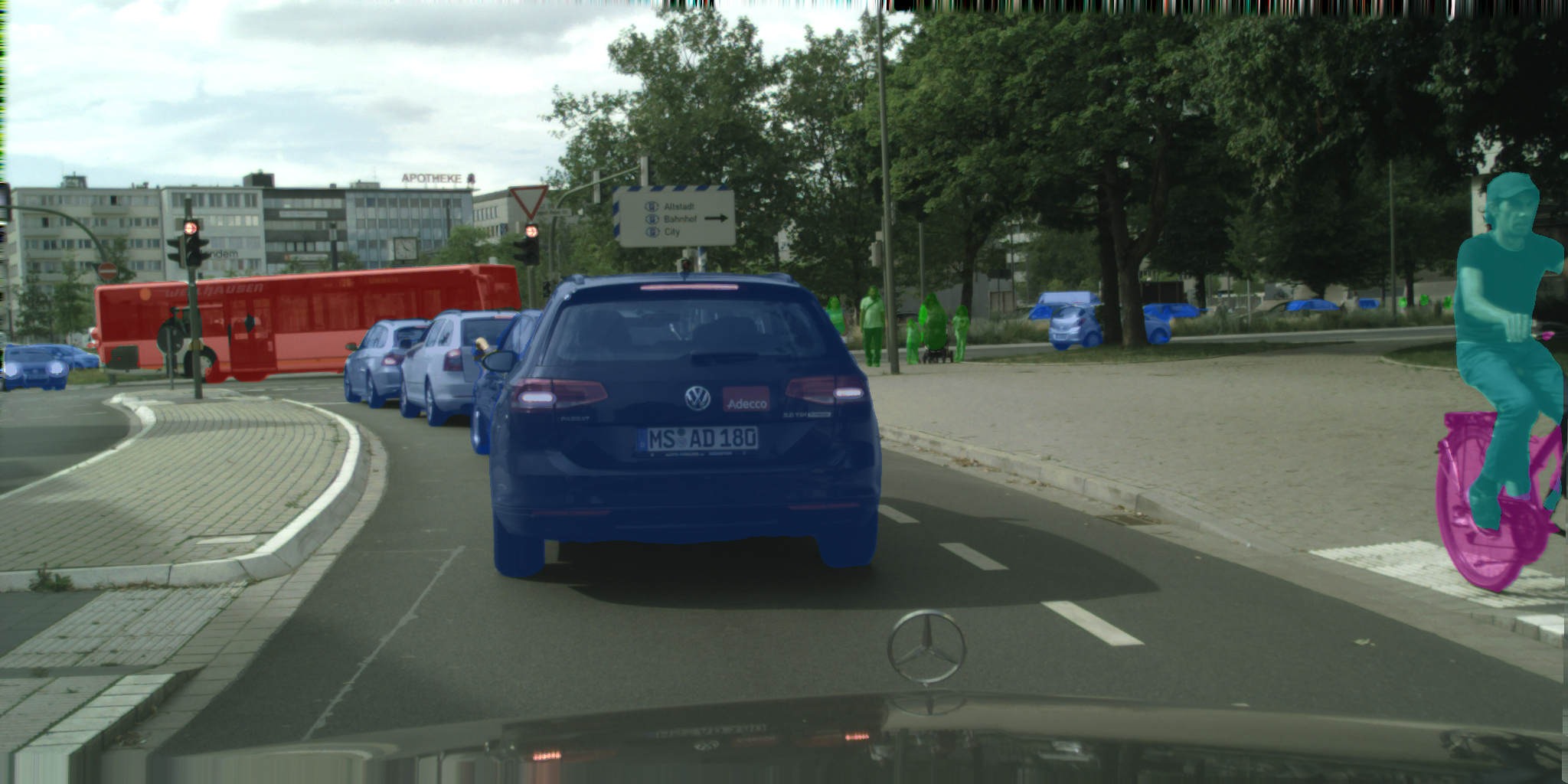}
    \caption{CompactInst (M\,$\nicefrac{1}{8}$\,PF\,$\nicefrac{1}{16}$)}
  \end{subfigure}
  \caption{%
    \textbf{Example~2} (1/2) - FastInst and CompactInst on a
    Cityscapes \texttt{test} image.%
  }
  \label{fig:qual-exp2a}
\end{figure}

\begin{figure}[p]
  \centering
  \begin{subfigure}[b]{\linewidth}
    \includegraphics[width=\linewidth]{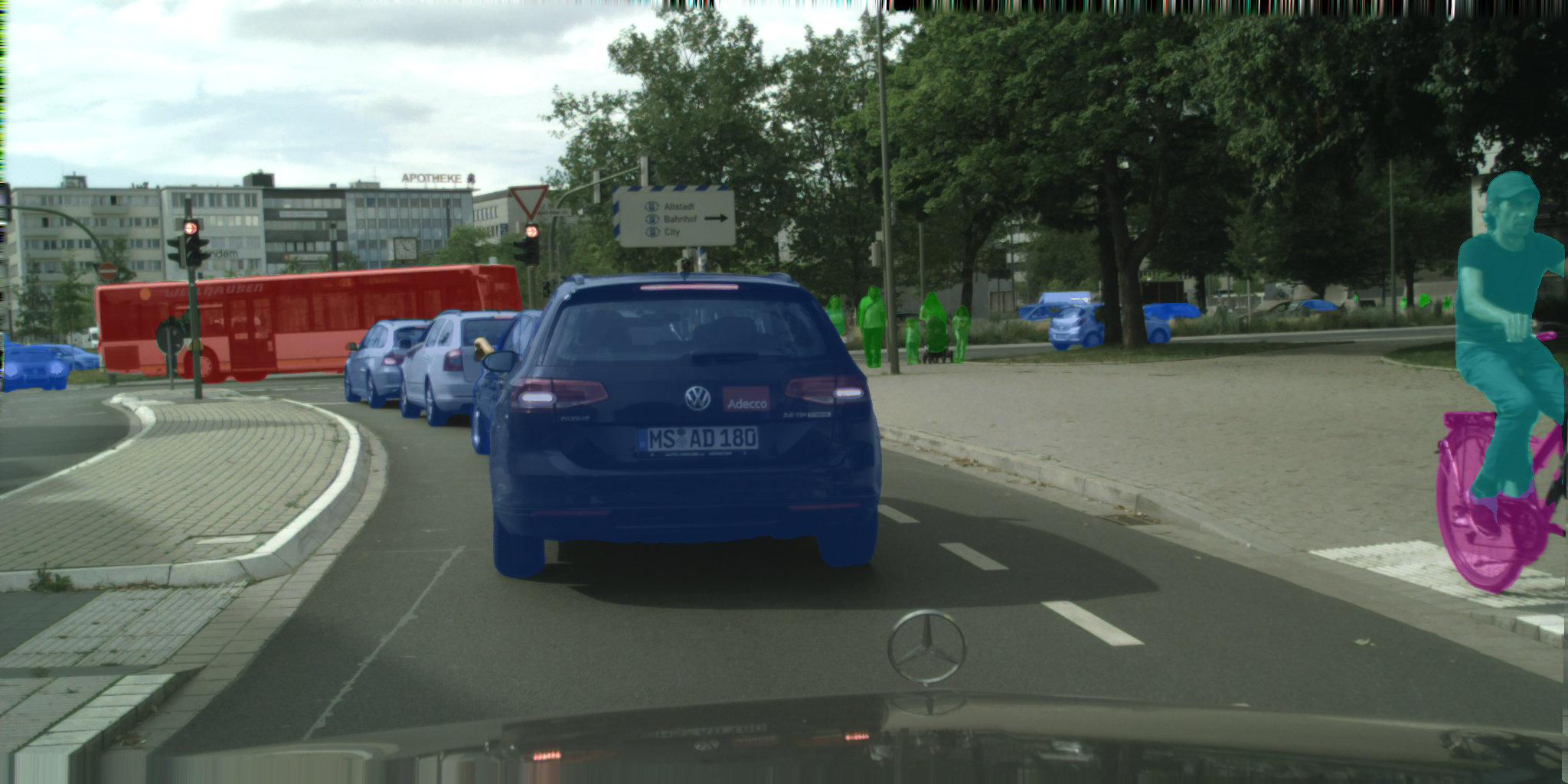}
    \caption{CompactInst w/ M\,$\nicefrac{1}{4}$\,PF\,$\nicefrac{1}{16}$}
  \end{subfigure}\\[4pt]
  \begin{subfigure}[b]{\linewidth}
    \includegraphics[width=\linewidth]{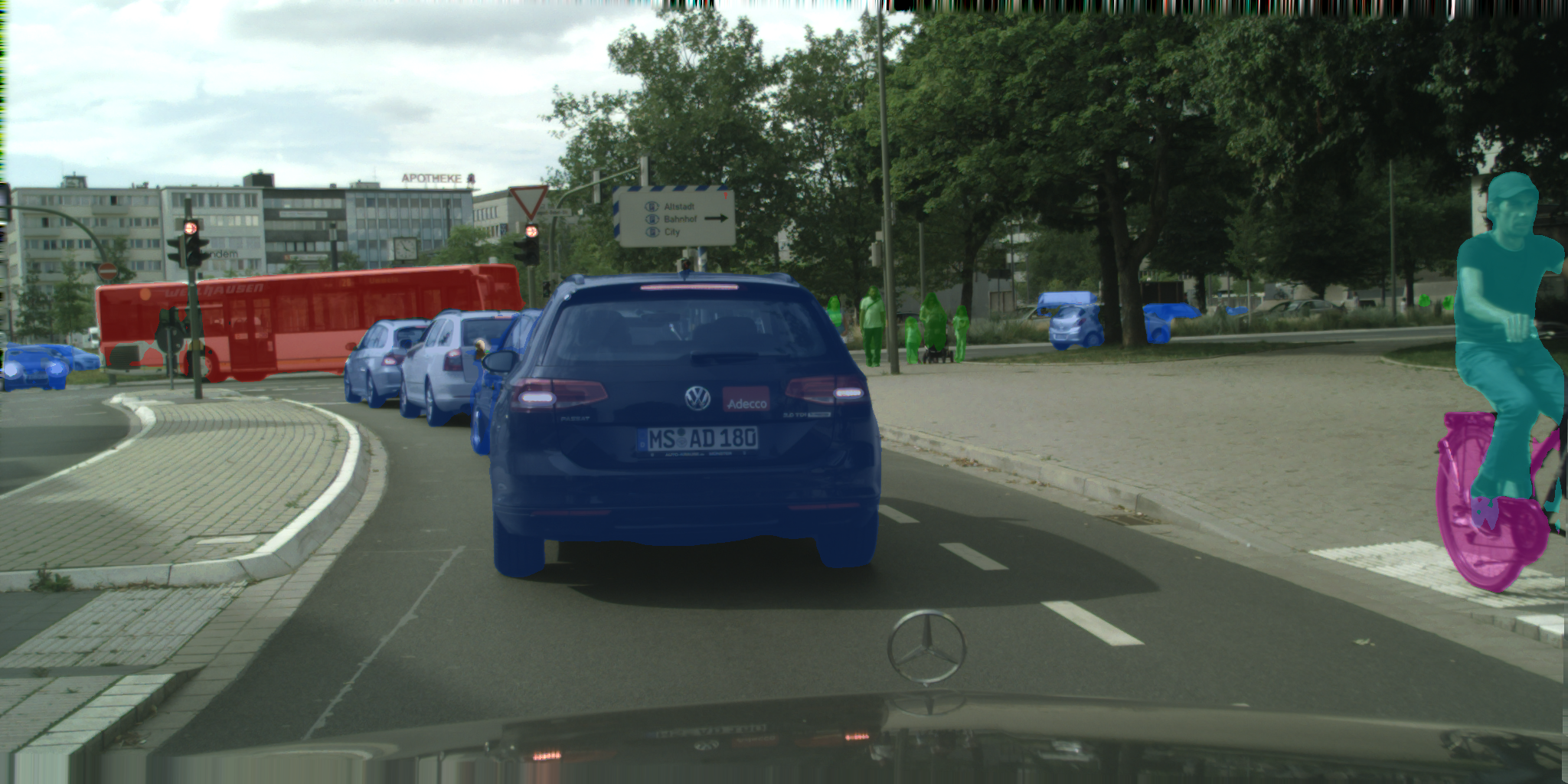}
    \caption{CompactInst w/ PF\,$\nicefrac{1}{32}$}
  \end{subfigure}
  \caption{%
    \textbf{Example~2} (2/2) - CompactInst variants on the same
    Cityscapes \texttt{test} image as \cref{fig:qual-exp2a}.%
  }
  \label{fig:qual-exp2b}
\end{figure}

\Cref{fig:qual-exp1a,fig:qual-exp1b,fig:qual-exp2a,fig:qual-exp2b}
show instance segmentation predictions of four models on two
Cityscapes \texttt{test} images: FastInst, CompactInst
(M\,$\nicefrac{1}{8}$\,PF\,$\nicefrac{1}{16}$), CompactInst with
M\,$\nicefrac{1}{4}$\,PF\,$\nicefrac{1}{16}$, and CompactInst with
PF\,$\nicefrac{1}{32}$.
Predicted instances are rendered as transparent masks with
class-specific colours:
\textcolor[rgb]{0.12,0.47,0.71}{\textbf{car}},
\textcolor[rgb]{0.17,0.63,0.17}{\textbf{person}},
\textcolor[rgb]{0.84,0.15,0.16}{\textbf{bus}},
\textcolor[rgb]{0.58,0.40,0.74}{\textbf{bicycle}},
\textcolor[rgb]{0.09,0.75,0.81}{\textbf{rider}}.
This colouring enables quick assessment of class predictions; however,
overlapping instances of the same class can be harder to distinguish
because their masks share a similar colour.

\clearpage
\paragraph{Analysis.}
FastInst and the default CompactInst produce relatively similar
predictions, with mostly minor mask-boundary differences - consistent
with their comparable quantitative performance on Cityscapes
(Table~2 in the main paper).

The M\,$\nicefrac{1}{4}$ variant shows noticeably better small-object
segmentation.
In Example~1 (\cref{fig:qual-exp1b}a), it segments the legs of the
small person on the road more accurately than the other models and
also avoids misclassifying the car in the back left as a person.
In Example~2 (\cref{fig:qual-exp2b}a), it is the only model that
detects all four extremely small and heavily occluded persons on the
right side of the scene, segmenting them with reasonable accuracy.
These observations visually confirm the strong AP$_S$ gains reported
for M\,$\nicefrac{1}{4}$ in the main paper ($+2.0$\,AP$_S$, $+35\,\%$
relative on Cityscapes \texttt{val}).

The PF\,$\nicefrac{1}{32}$ variant, which applies the most aggressive
token reduction, shows visible degradation on small instances.
In Example~1 (\cref{fig:qual-exp1b}b), it misses the person on the
right that all other models detect.
In Example~2 (\cref{fig:qual-exp2b}b), it fails to detect two of the
four small occluded persons on the right as well as a nearby
car - consistent with the AP drop observed for this variant in the
quantitative evaluation.

\section{TensorRT Reproducibility}
\label{sec:supp-trt}

All latency measurements use batch-1 static TensorRT engines built from
ONNX at opset~17 with the FP16 builder flag and no custom plugins. Each
resolution has a separate optimization profile with
$\text{min}=\text{opt}=\text{max}$. For ONNX export, CompactInst's
\texttt{adaptive\_max\_pool2d} is replaced by an equivalent fixed
\texttt{max\_pool2d} whose kernel and stride equal the known pooling
factor. TensorRT versions were 10.4.0 on the RTX~A4000, 8.6.2 on
AGX~Orin, and 8.5.2.2 on Orin~Nano.

Decoder-only engines receive preallocated synthetic feature tensors that
match the pixel-decoder output shapes. CUDA events measure engine execution;
allocation, data transfer, engine loading, and post-processing are excluded.
The A4000 validation uses its stock 140\,W limit, requested GPU/memory clocks
of 1.56/7.0\,GHz, and medians of five alternating trials with 50 warmup and
200 timed iterations. AGX~Orin measurements use the platform's
maximum-performance clock configuration. The matched Mask2Former experiment
uses an Orin~Nano 8\,GB with JetPack 5.1.1 (Jetson Linux R35.3.1) in 15\,W
mode, with the GPU and EMC fixed at 624.75 and 2133\,MHz. We report medians
of five randomized trials.

\section{ResNet-18 Cityscapes Accuracy}
\label{sec:supp-r18-acc}

The main paper reports accuracy using ResNet-50, while model-level
latency uses ResNet-18 because the ResNet-50 TensorRT engines exceed
Orin~Nano's memory at high resolutions. As an additional consistency
check, \cref{tab:supp-r18-acc} reports Cityscapes accuracy for the
corresponding ResNet-18 models.

\begin{table}[H]
\centering
\small
\setlength{\tabcolsep}{6pt}
\renewcommand{\arraystretch}{1.2}
\begin{tabular}{@{}l cccc@{}}
\hline
 & \multicolumn{2}{c}{\texttt{test}} &
   \multicolumn{2}{c}{\texttt{val}} \\
\cline{2-3} \cline{4-5}
Method (ResNet-18) & AP & AP$_{50}$ & AP & AP$_{50}$ \\
\hline
FastInst
  & 23.4 & 44.9 & 26.0 & 48.6 \\
CompactInst (M\,$\nicefrac{1}{8}$\,PF\,$\nicefrac{1}{16}$)
  & 23.6 & 44.1 & 26.2 & 47.8 \\
CompactInst (M\,$\nicefrac{1}{4}$\,PF\,$\nicefrac{1}{16}$)
  & 25.1 & 46.8 & 26.7 & 47.6 \\
\hline
\end{tabular}
\caption{Cityscapes instance-segmentation accuracy of the ResNet-18
  models used for the model-latency measurements.}
\label{tab:supp-r18-acc}
\end{table}

\clearpage
\section{Post-Processing Latency}
\label{sec:supp-postproc}

\begin{figure}[H]
  \centering
  \includegraphics[width=0.92\linewidth]{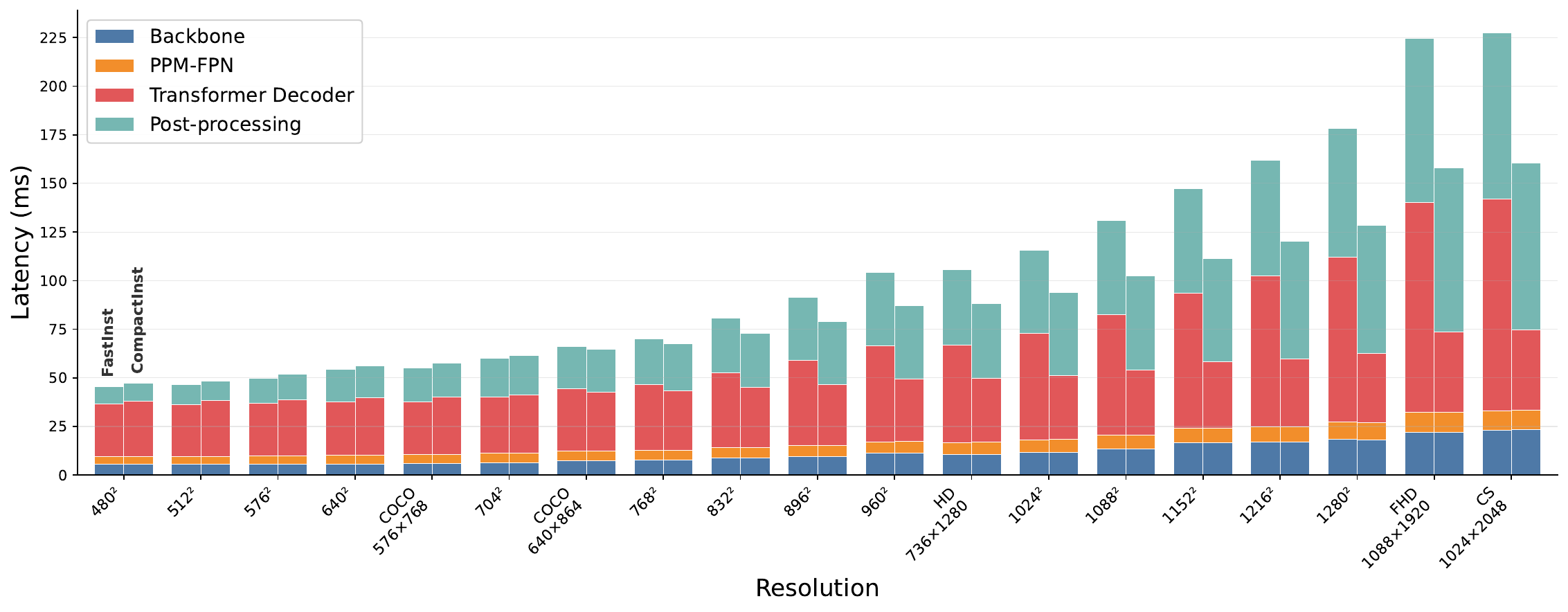}
  \caption{Per-component AMP latency on Jetson AGX~Orin with a
    ResNet-18 backbone (batch size~1).}
  \label{fig:supp-breakdown-r18}
\end{figure}

\begin{figure}[H]
  \centering
  \includegraphics[width=0.92\linewidth]{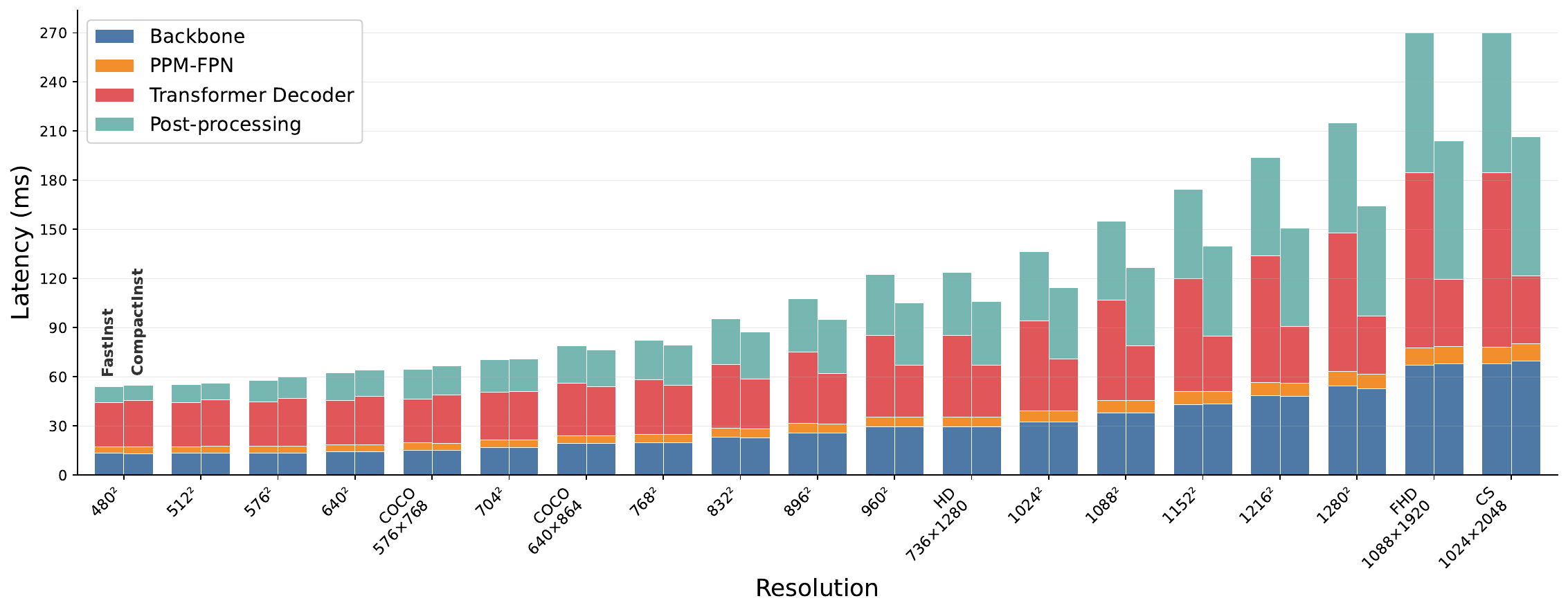}
  \caption{Per-component AMP latency on Jetson AGX~Orin with a
    ResNet-50 backbone (batch size~1).}
  \label{fig:supp-breakdown-r50}
\end{figure}

We measure component-level PyTorch AMP latency at batch size~1 on
Jetson AGX~Orin across 19 input resolutions. The breakdown separately
times the backbone, pixel decoder, transformer decoder, and
post-processing for ResNet-18 and ResNet-50
(\cref{fig:supp-breakdown-r18,fig:supp-breakdown-r50}). At Cityscapes
resolution, post-processing takes approximately 85\,ms and is the
second-largest FastInst component and the largest CompactInst
component with either backbone.

Post-processing upsamples the predicted masks to the output resolution
and applies elementwise thresholding and mask scoring. These operations
move full-resolution mask tensors while performing little arithmetic.
Its approximately linear growth with pixel count is therefore
consistent with a separate memory-bandwidth bottleneck outside the
transformer decoder targeted by CompactInst.